\documentclass[useAMS,usenatbib]{mnras}

\usepackage{graphicx}
\usepackage{enumerate}
\usepackage{xcolor} % use with \textcolor
\usepackage{amsmath}
\newcommand{\logoh}{12$+$log(O/H)}

\newcommand{\msun}{$M_\odot$}

\newcommand{\htwo}{H$_2$}
\newcommand{\hi}{H{\sc i}}

\newcommand{\mstar}{M$_{*}$}

\newcommand{\epsfinv}{$\epsilon_{*}^{-1}$}
\newcommand{\epsf}{$\epsilon_{*}$}
\newcommand{\epff}{$\epsilon_{\rm ff}$}
\newcommand{\tff}{$t_{\rm ff}$}
\newcommand{\esca}{$e_{\rm scale}$}
\newcommand{\taud}{$\tau_{\rm depl}$}
\newcommand{\mg}{$M_{g}$}

\title[Metallicity and SFR coevolution]{Coevolution of metallicity and star formation 
in galaxies to $z\simeq 3.7$. II. A theoretical model}
\author[Leslie Hunt, Pratika Dayal, Laura Magrini, Andrea Ferrara]{Leslie Hunt$^{1}$\thanks{E-mail: hunt@arcetri.astro.it}, Pratika Dayal$^{2}$, 
Laura Magrini$^{1}$ and Andrea Ferrara$^{3}$ \\
$^{1}$INAF/Osservatorio Astrofisico di Arcetri, Largo Enrico Fermi 5, 50125 Firenze, Italy\\
$^{2}$Kapteyn Astronomical Institute, University of Groningen, P.O. Box 800,
9700 AV Groningen, The Netherlands\\
$^{3}$Scuola Normale Superiore, Piazza dei Cavalieri 7, I-56126 Pisa, Italy}
\begin{document}

\date{draft version 8 August 2016}

\pagerange{\pageref{firstpage}--\pageref{lastpage}} \pubyear{2016}

\maketitle

\label{firstpage}

\begin{abstract}
Recent work suggests that galaxy evolution, and the
build-up of stellar mass (\mstar) over cosmic time, is characterized
by changes with redshift of star formation rate (SFR) and oxygen abundance (O/H).
In a companion paper,  
we have compiled a large dataset to study Metallicity Evolution
and Galaxy Assembly (MEGA), consisting
of $\sim$1000 galaxies to $z\simeq 3.7$ with a common O/H calibration. 
Here we interpret the MEGA scaling relations of \mstar, SFR, and O/H
with an updated version of the model presented by \citet{dayal13}. 
This model successfully
reproduces the observed O/H ratio of $\sim$80\,000 galaxies selected
from the Sloan Digital Sky Survey to within 0.05$-$0.06\,dex. 
By extending the model to the higher redshift MEGA sample, 
we find that although the specific mass
loading of outflows does not change measurably during the evolution, the
accretion rate and gas content of galaxies increase significantly with
redshift. These two effects can 
explain, either separately or possibly in tandem, the observed lower metal abundance
of high-$z$ galaxies.
\end{abstract}


\section{Introduction}
\label{sec:intro}

Galaxy evolution takes place through the build-up over time of
stellar mass (\mstar) through various episodes of star formation.
As stars evolve, metals are produced, so that metallicity
and its relation with \mstar\ and star-formation rate (SFR) are
important gauges of star-formation history. 
Thus, metal content, typically measured through
the gas-phase oxygen abundance (O/H), the most abundant heavy element produced by
massive stars, can be used to assess a galaxy's evolutionary state.
The integrated star-formation activity over the lifetime of the galaxy is
measured by stellar mass, while the star-formation rate (SFR) measures how
much gas is being converted into stars over the time scale of the SFR indicator.
The gas-phase metallicity ($Z$) quantifies the production of metals from high-mass stars,
but also indicates the degree to which galaxies interact with their environment through 
gas accretion and outflows in the form of galactic winds.

Several well-established scaling relations reflect the intimate link between
\mstar, SFR, and O/H.
A mass-metallicity relation (MZR) connecting stellar mass and $Z$ is clearly present
in the Local Universe \citep[e.g.,][]{tremonti04}, and it apparently extends to
the highest redshifts examined so far
\citep[e.g.,][]{erb06a,maiolino08,mannucci09,cresci12,xia12,yabe12,henry13,cullen14,zahid14,troncoso14,steidel14,wuyts14,ly15,delosreyes15,onodera16}.
Stellar mass and SFR are also related in a SF ``main sequence'' (SFMS)
both locally \citep{brinchmann04,salim07} and at high redshift
\citep[e.g.,][]{noeske07,elbaz11,karim11,wuyts11,speagle14}. 
It is generally agreed that
the high-redshift relations of both the MZR and the SFMS are similar in
form to the local ones, but they show different normalizations: 
at a given \mstar, SFR (and sSFR) increases with increasing redshift
while metallicity decreases.
Moreover, both relations show an inflection at high \mstar, with
a decrease in O/H and SFR
\citep{tremonti04,wyder07,noeske07,whitaker14,lee15,gavazzi15}.

In order to observationally constrain the evolution of metallicity with
redshift, we have compiled a new dataset of $\sim 1000$ star-forming galaxies 
from $z\simeq 0$ to $z\sim3.7$ with nebular oxygen abundance measurements;
we have dubbed this compilation the ``MEGA'' dataset, corresponding to 
{\it Metallicity Evolution and Galaxy Assembly}.
This dataset presents a marked improvement over the dataset used by \citet{hunt12}
because of the inclusion of several more high$-z$ samples 
and, more importantly, because of a common metallicity calibration.
The coevolution of SFR and O/H with redshift in the MEGA dataset
is presented in  a companion paper (Hunt et al. 2016, hereafter Paper I), and summarized in Section \ref{sec:coevo}.
Here, we describe a physical model for understanding the coevolution of \mstar, SFR,
and O/H.
This is an extension of the model presented by \citet{dayal13},
and also a re-assessment of the model parameters with different O/H calibrations,
different stellar yields, and a different Initial Mass Function (IMF).
Section \ref{sec:theory} describes this analytical approach 
%to a physical understanding of the scaling relations with redshift,
to physically interpret the scaling relations locally,
and Section \ref{sec:fmrhighz} extends the analytical formulation to quantify evolution of the scaling relations with redshift.
We discuss the results of our modeling in Sect. \ref{sec:discussion}, and 
summarize our conclusions in Sect. \ref{sec:conclusions}.
As in Paper I, we use a \citet{chabrier03} IMF throughout.
%and, when necessary, have adopted the conversions for \mstar\ and SFR given by \citet{speagle14}.

\section{Metallicity and SFR coevolution}
\label{sec:coevo}

Paper I presents common O/H calibrations for 990 galaxies from $z\sim0$ to $z\sim3.7$
culled from 19 different samples.
The MEGA dataset spans a range of $10^5$ in \mstar, $10^6$ in SFR, and almost
two orders of magnitude in O/H.
%The \mstar\ and SFR calculations rely on the \citet{chabrier03} IMF;
The conversion to common metallicity calibrations was performed according to the transformations given by
\citet{kewley08}.
We chose three O/H calibrations, because of their similarity to electron-temperature
or ``direct'' methods \citep[e.g.,][]{andrews13}:
\citet[][hereafter D02]{denicolo02}
and the nitrogen and oxygen$+$nitrogen-based abundances by \citet[][hereafter PP04N2 and PP04O3N2, respectively]{pettini04}.

We hypothesized that the inflections in the MZR and SFMS at high \mstar\ were compensating
one another, and also that the {\it decrease} in O/H at high redshift was somehow quantitatively
associated with the {\it increase} in SFR.
This hypothesis was essentially aimed at
assessing the possibility that the mutual correlations among \mstar,
SFR, and O/H could be expressed as a planar relation, i.e., with only two of the three
pseudo-observables being required to describe the variation.
From a Principal Component Analysis (PCA), we showed that this was indeed possible,
and found a 2-dimensional relation, namely the Fundamental Plane of
Metallicity (FPZ):

\begin{equation}
12+\log(O/H) = -0.14\,{\rm log(SFR)} + 0.37\,{\rm log(M_*)} + 4.82 \ \ 
\label{eqn:fpzall}
\end{equation}

\noindent
The FPZ is an accurate reflection of the relations among \mstar, SFR, and O/H at least to
within 0.16\,dex in O/H \citep[Eqn. (\ref{eqn:fpzall}) represents the best-behaved O/H calibration, PP04N2,][]{pettini04}.
As described in Paper I,
such a level of residuals is comparable to trends found in other galaxy samples
but with a smaller range in \mstar, SFR, and O/H.

We also analyzed whether or not the FPZ varied over time, and found a significant
correlation of the residuals with redshift, but with a residual standard error  of 0.16\,dex, the same as the 0.16\,dex uncertainty of the FPZ itself.
Thus, we concluded that the FPZ is approximately redshift invariant,
since any redshift variation of the FPZ is within the noise of the current data.

As a comparison to the MEGA dataset, here and in Paper I we also analyze the emission-line galaxies from the Sloan Digital Sky Survey (SDSS)
defined by \citet[][hereafter the SDSS10 sample]{mannucci10}.
Although its parameter-space coverage is relatively limited,
because of its large size this sample adds superb statistical power to our analysis, and also enables a comparison
with \citet{dayal13} who analyzed the same galaxies.
As for the MEGA dataset, we have converted the \citep[][KD02]{kewley02} metallicities reported by \citet{mannucci10} 
to the D02, PP04N2, and PP04O3N2 calibrations, as described in Paper I.

\section{Coevolution: physical insights}
\label{sec:theory}

Much theoretical effort has been devoted to understand the evolution
of the MZR and its dependence, if any, on SFR and the surrounding environment
\citep[e.g.,][]{peng10,peeples11,dave11,yates12,hopkins12,krumholzdekel12,dayal13,lilly13,schaye15}.
%but most attempts depend on assumptions about the relation
%between gas inflow from the surrounding intergalactic medium,
%gas outflow through galactic-scale winds
Here, following the analytical formalism of \citet{dayal13}, we present a physical basis for
the observed coevolution of SFR and O/H.
With a nebular oxygen abundance $X\equiv M_O/M_g$ where
$M_O$ and $M_g$ are the galaxy oxygen and total gas mass, respectively, 
SFR, O/H, and \mstar\ are related through a set of evolutionary
differential equations 
%we define
%a simple set of evolutionary ordinary differential equations
%\citep[see Eq. (1-3),][]{dayal13}.
\citep[see][]{dayal13}:

\begin{equation}
\frac{{\rm d}M_{\rm star}}{{\rm d}t}\,\equiv\,\psi\,=\,\epsilon_*\,M_g
\label{eqn:sfr}
\end{equation}

\begin{equation}
\frac{{\rm d}M_g}{{\rm d}t}\,=\,-(1-R) \psi + (a-w) \psi
\label{eqn:gas}
\end{equation}

\begin{equation}
M_g\,\frac{{\rm d}X}{{\rm d}t}\,=\,y(1-R)\psi - a X \psi
\label{eqn:metals}
\end{equation}

\noindent
Here \epsfinv\ is the star-formation timescale and we have renamed SFR as $\psi$;
we assume that the SFR is proportional to the gas mass, and
that the infalling gas is much less metal enriched than
the ambient ISM, $X_i\,\ll\,X$.
The two constants, $R$ and $y$, represent the returned fraction from stars
and the yield per stellar generation, respectively, and depend on the IMF; we
%have adopted $(R,y)\,=\,(0.79, 0.0871)$ which are consistent with
%a Salpeter IMF for a lower (upper) mass limit of 1 (100)\,\msun.
have adopted $(R,y)\,=\,(0.46, 0.05)$, 
taken from \citet{vincenzo16} using the stellar yields from \citet{nomoto13}
assuming a \citet{chabrier03} IMF\footnote{The stellar yields from \citet{romano10}
are similar, with $(R,y)\,=\,(0.45,0.06)$. Our values for $(R,y)$ are roughly
the mean values of the yields over the range of metallicities given by \citet{vincenzo16} since our model
does not contemplate time variation of the yields.}. 
The lower (upper) mass limit for these yields is 0.1\,\msun\ (100\,\msun).
%This is a potential problem for our use of a Chabrier IMF, 
%since although the upper limits of 100\,\msun\ are the same for both IMFs,
%the yields change with the fraction of the total stellar mass residing
%in stars more massive than $\sim$8\,\msun\ \citep[e.g.,][]{romano05,romano10,vincenzo16}.
%Nevertheless, to be consistent with the older version of the \citet{dayal13}
%models, we have maintained these values;
%ultimately it is a question of normalization and does not change our conclusions.

Like \citet{dayal13}, 
we further assume that gas outflow through
winds, $w(M)\psi$ is proportional to the SFR, where $M$ is the total (dark$+$baryonic)
galaxy mass.
There is some observational evidence of such a proportionality,
at least in the hot and/or ionized gas components of galaxies
\citep[e.g.,][]{martin05,veilleux05}. 
For convenience in the analytical formulation, we also take gas
accretion, $a(M)\psi$, to be proportional to SFR.
Although there is little evidence to directly support such
a proportionality, it may be that gas is replenished in galaxies
by cooling of pre-existing ionized gas in expanding supershells or galactic fountains  
\citep[e.g.,][]{hopkins08,fraternali08,fraternali12}.
%However, we make no claim of causality in our assumption of proportionality between
%gas accretion and SFR, but rather, as in \citet{dayal13}, exploit its convenience for the analytical formulation
%of the model.

The solution of these equations can be written as:

\begin{equation}
%X = \frac{y(1-R)}{a} \left[ 1 - \left( \frac{M_g}{M_{g0}} \right)^{-\alpha} \right] \nonumber \\
X = \frac{y(1-R)}{a} \left( 1 - \mu^{-\alpha} \right) 
\label{eqn:X}
\end{equation}

\noindent
% for making tex images for talk
%where $\mu\,\equiv\,\frac{M_g}{M_{g0}}\,=\,\psi [\epsilon_* M_{g0}]^{-1}$\ \ ,
where $\mu\,\equiv\,{M_g}/{M_{g0}}\,=\,\psi [\epsilon_* M_{g0}]^{-1}$,
and 
%\medskip
%$\alpha\equiv a\,(R-1+a-w)^{-1}$\ \ .
$\alpha\equiv a\,(R-1+a-w)^{-1}$.
%\medskip
The initial gas mass can be expressed as: 
\begin{equation}
M_{g0}\,\simeq\,M_{\rm star}\,(1 + w -a) + M_g\,=\,(\Omega_b/\Omega_M)\,M
\label{eqn:mgas}
\end{equation}
Altogether, there are five free parameters in the ($z \simeq 0$) model: the inverse of
the star-formation timescale, \epsf; and the coefficients and power-law indices describing
the mass dependence of infall through gas accretion ($a$), 
$a(M_{\rm star})\,=\,a_{\rm coeff}\,(M_{\rm star}/M_0)^{a_{\rm pow}}$;
and of outflow by galactic winds ($w$),
$w(M_{\rm star})\,=\,w_{\rm coeff}\,(M_{\rm star}/M_0)^{w_{\rm pow}}$.

The solution to these equations results in an implicit star-formation history
(SFH) assumed by our models, namely an exponentially declining one.
Such a SFH is observationally supported by some studies
\citep[e.g.,][]{eyles07,stark09}, but contested by others
\citep[e.g.,][]{pacifici15,salmon15}.
In our case, the exponential time scale depends on halo mass (here assumed
to be proportional to \mstar).
As described by \citet{dayal13}, this means that 
galaxies of different masses build up their metal contents at different rates: 
galaxies start from low $X$, low \mstar, and high $\psi$ values and then move towards higher metallicities as their \mstar\ increases. 
However, depending on their mass, they move at different velocities along the track: 
the most massive observed galaxies are very evolved with a large specific age (i.e., in units of the star formation timescale) 
and move essentially along constant metallicity curves; smaller objects are younger 
and are still gently building up their metal content.  

\begin{figure*}
\vspace{\baselineskip}
\hbox{
% \includegraphics[width=0.95\textwidth]{../data/newfmr_sdss_d02_ppn2_ppo3_theo_newcolors.png}
% 16/5/2016
 \includegraphics[width=0.95\textwidth]{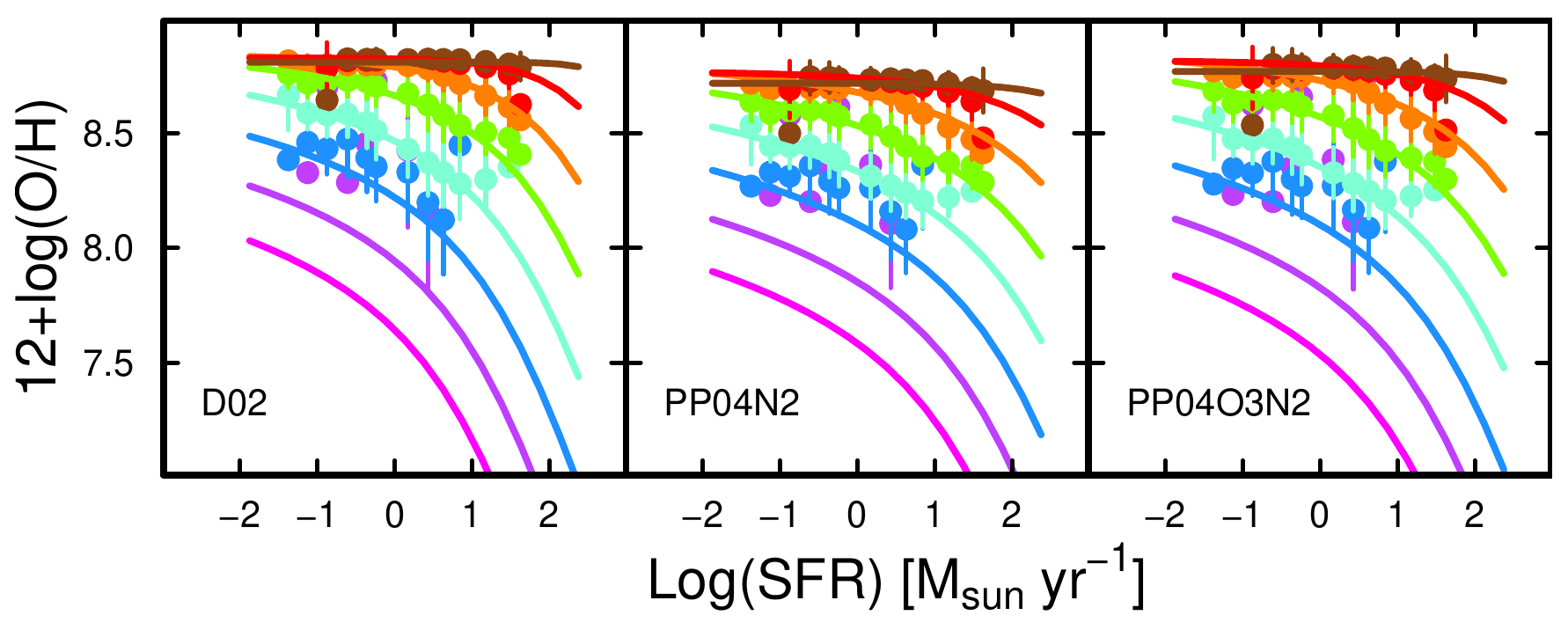}
}
\caption{SDSS10 galaxies: \logoh\ plotted against (log of) SFR; the data points
correspond to the average \logoh\ (with the three O/H calibrations)
for SDSS galaxies in a given SFR bin, and
the error bars to the $1\sigma$ spread of the data. 
The curves show three FPZ models with different O/H calibrations 
(D02, PP04N2, PP04O3N2) as discussed in the text.
The color coding is by (log) \mstar:
%7-7.5 (magenta), 
7.5-8 (magenta), 8-8.5 (purple),
8.5-9 (blue), 9-9.5 (cyan), 9.5-10 (green), 10-10.5 (orange), 10.5-11 (red), 11-11.5 (brown). 
The model and the data are in excellent agreement with a residual scatter of 0.05$-$0.07\,dex,
according to the O/H calibration (see Table \ref{tab:models}).
}
\label{fig:fmrsdss}
\end{figure*}

\begin{figure*}
\vspace{\baselineskip}
\hbox{
% 16/5/2016, 24/6/2016
 \includegraphics[width=0.95\textwidth]{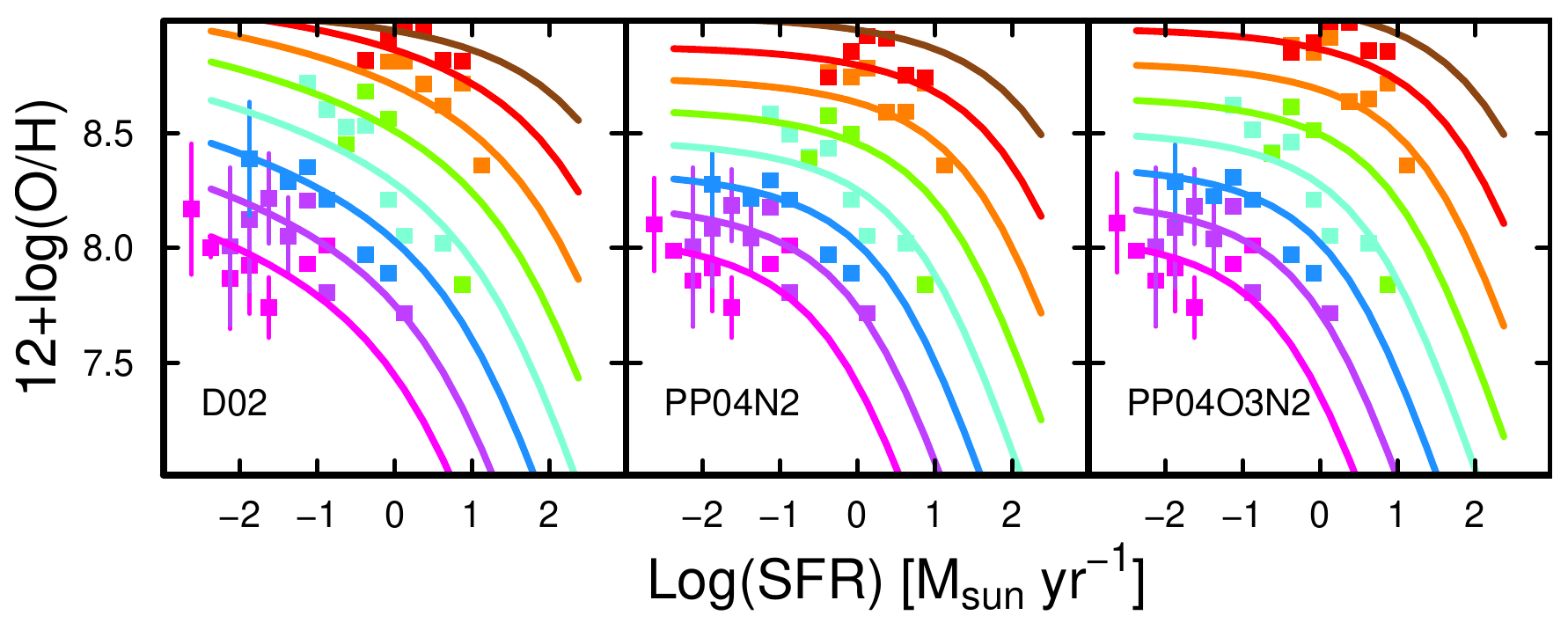}
}
\caption{MEGA galaxies: \logoh\ plotted against (log of) SFR; the data points
correspond to the average \logoh\ (with the three O/H calibrations)
for galaxies in a given SFR bin, and
the error bars to the $1\sigma$ spread of the data. 
The curves show three FPZ models with different O/H calibrations 
(D02, PP04N2, PP04O3N2) as discussed in the text; here we show the FPZ best-fit MEGA model,
rather than the FPZ best-fit SDSS model given in Fig. \ref{fig:fmrsdss}. 
As in Fig. \ref{fig:fmrsdss}, the color coding is by (log) \mstar:
%7-7.5 (magenta), 
7.5-8 (magenta), 8-8.5 (purple),
8.5-9 (blue), 9-9.5 (cyan), 9.5-10 (green), 10-10.5 (orange), 10.5-11 (red), 11-11.5 (brown). 
Only bins with more than 2 galaxies are shown; in some cases the standard deviations
within the bins are smaller than the data point.
The residual scatter of the model is 0.24$-$0.26\,dex,
according to the O/H calibration (see Table \ref{tab:models}).
}
\label{fig:fmrmega}
\end{figure*}

We can accommodate this model (hereafter the FPZ model) to the MEGA dataset and the SDSS10 sample for the three O/H calibrations
considered here by fitting for these five parameters.
Fig. \ref{fig:fmrsdss} illustrates the resulting best-fit FPZ model overplotted
on the binned SDSS10 data for all three O/H calibrations (D02, PP04N2, PP04O3N2).
Since this is a local calibration,
the FPZ model fit to the MEGA dataset included only galaxies with $z\leq 0.1$;
Fig. \ref{fig:fmrmega} shows this best fit to the MEGA data set of the FPZ model.
Table \ref{tab:models} reports the best-fit parameters.

The model is an excellent representation of the SDSS10 data
\citep[see also][]{dayal13}; over $\sim$80\,000 individual data points,
the best-fit model with only five free parameters gives a residual error
in \logoh\ predicted vs. observed of 0.05--0.07\,dex, according to the O/H calibration. 
The model is somewhat less successful at fitting the MEGA dataset, with mean O/H residuals
of predicted vs. observed of $\sim 0.24-0.26$\,dex, larger than the FPZ residuals of $\sim 0.16$\,dex
(see Paper I).
However, these galaxies are only a subset of the entire MEGA sample (because of the redshift limitation),
and the best-fit MEGA parameters are generally consistent, within the errors, to SDSS (see below
for more details).

%%%%%%%%%%%%%%%%%%%%%%%%
\begin{table*} 
\setlength{\tabcolsep}{3pt}
\begin{center} 
\caption {FPZ model best-fit parameters$^{\mathrm a}$}
{\scriptsize
\begin{tabular}{lrcccccccccc}
\hline 
\multicolumn{1}{c}{Sample} &
\multicolumn{1}{c}{Degrees of} &
\multicolumn{1}{c}{$\sigma_{\rm fit}^{\mathrm b}$} &
\multicolumn{1}{c}{O/H} &
\multicolumn{1}{c}{$a_{\rm coeff}$} &
\multicolumn{1}{c}{$a_{\rm pow}$} &
\multicolumn{1}{c}{$w_{\rm coeff}$} &
\multicolumn{1}{c}{$w_{\rm pow}$} &
\multicolumn{1}{c}{\epsf} \\
& \multicolumn{1}{c}{freedom} & &\multicolumn{1}{c}{calibration$^{\mathrm c}$} &
&&&& (Gyr$^{-1}$)\\
\multicolumn{1}{c}{(1)} &
\multicolumn{1}{c}{(2)} &
\multicolumn{1}{c}{(3)} &
\multicolumn{1}{c}{(4)} &
\multicolumn{1}{c}{(5)} &
\multicolumn{1}{c}{(6)} &
\multicolumn{1}{c}{(7)} &
\multicolumn{1}{c}{(8)} &
\multicolumn{1}{c}{(9)} &
\multicolumn{1}{c}{(10)} &
\multicolumn{1}{c}{(11)} \\
\hline 
\multicolumn{11}{c}{Local Universe} \\
\\
% 16/5/2016 new values
% see ~/statistics/mzr_onodera/mzrnew_NEWfmrpredict_z=0_R.txt for fits
MEGA$^{\mathrm d}$ & 252 & 0.264 & D02    & $0.47\,\pm\,0.45$ & ~~$0.22\pm0.60$  & $15.55\,\pm\,4.0$ & $-0.34\pm0.04$   & $2.0\,\pm\,1.3$ & $-$ & $-$ \\ 
MEGA$^{\mathrm d}$ & 252 & 0.243 & PP04N2 & $0.88\,\pm\,0.32$ & $-0.27\pm0.06$ & $9.9\,\pm\,4.7$  & $-0.29\pm0.06$   & $1.1\,\pm\,0.8$ & $-$ & $-$ \\ 
MEGA$^{\mathrm d}$ & 252 & 0.251 & PP04O3N2 & $0.73\,\pm\,0.29$ & $-0.30\pm0.06$ & $8.6\,\pm\,3.9$  & $-0.32\pm0.06$   & $0.9\,\pm\,0.6$ & $-$ & $-$ \\ 
\\
SDSS10    & 78536 & 0.052 & D02      & $0.994\,\pm\,0.001$ & $0.036\pm0.002$ & $11.25\,\pm\,0.11$ & $-0.449\pm0.002$ & ~$6.3\,\pm\,0.2$ & $-$ & $-$ \\ 
SDSS10    & 78536 & 0.063 & PP04N2   & $1.136\,\pm\,0.005$ & $0.104\pm0.005$ & $20.55\,\pm\,0.26$ & $-0.413\pm0.003$ & $13.1\,\pm\,0.8$ & $-$ & $-$ \\ 
SDSS10    & 78536 & 0.068 & PP04O3N2 & $1.020\,\pm\,0.004$ & $0.096\pm0.005$ & $18.09\,\pm\,0.21$ & $-0.451\pm0.003$ & ~$8.2\,\pm\,0.4$ & $-$ & $-$ \\ 
\\
\hline 
\multicolumn{11}{c}{$z>0$: $\epsilon_{*}(z)\,=\epsilon_{*}(0)\,({\rm M}_{\rm star}/{10^{9.5}})^{\gamma(\epsilon_*)} \ (1+z)^{\delta(\epsilon_*)}$} \\
 & & & & & & & & & \multicolumn{1}{c}{$\gamma(\epsilon_*)$} & \multicolumn{1}{c}{$\delta(\epsilon_*)$} \\
\multicolumn{1}{c}{(1)} &
\multicolumn{1}{c}{(2)} &
\multicolumn{1}{c}{(3)} &
\multicolumn{1}{c}{(4)} &
\multicolumn{1}{c}{(5)} &
\multicolumn{1}{c}{(6)} &
\multicolumn{1}{c}{(7)} &
\multicolumn{1}{c}{(8)} &
\multicolumn{1}{c}{(9)} &
\multicolumn{1}{c}{(10)} &
\multicolumn{1}{c}{(11)} \\
% 27/6/2016 redone 
MEGA$^{\mathrm e}$      & 25 & 0.060 & 
D02  & $0.994\,\pm\,0.001$ & $0.036\pm0.002$ & $11.25\,\pm\,0.11$ & $-0.449\pm0.002$ & ~$6.3\,\pm\,0.2$ 
& $-0.23\pm0.07$ & $-0.16\pm0.09$ \\ 
MEGA$^{\mathrm e}$      & 25 & 0.038 &
PP04N2   & $1.136\,\pm\,0.005$ & $0.104\pm0.005$ & $20.55\,\pm\,0.26$ & $-0.413\pm0.003$ & $13.1\,\pm\,0.8$ 
& $-0.49\pm0.05$ & $-0.17\pm0.08$ \\ 
MEGA$^{\mathrm e}$      & 25 & 0.046 & 
PP04O3N2 & $1.020\,\pm\,0.004$ & $0.096\pm0.005$ & $18.09\,\pm\,0.21$ & $-0.451\pm0.003$ & ~$8.2\,\pm\,0.4$ 
& $-0.44\pm0.05$ & $-0.10\pm0.08$ \\ 
\\
Schechter$^{\mathrm f}$ & 28 & 0.079 & 
D02      & $0.994\,\pm\,0.001$ & $0.036\pm0.002$ & $11.25\,\pm\,0.11$ & $-0.449\pm0.002$ & ~$6.3\,\pm\,0.2$ 
& $-0.40\pm0.06$ & $-0.68\pm0.10$ \\ 
Schechter$^{\mathrm f}$ & 28 & 0.025 & 
PP04N2   & $1.136\,\pm\,0.005$ & $0.104\pm0.005$ & $20.55\,\pm\,0.26$ & $-0.413\pm0.003$ & $13.1\,\pm\,0.8$ 
& $-0.54\pm0.03$ & $-0.23\pm0.05$ \\ 
Schechter$^{\mathrm f}$ & 28 & 0.025 & 
PP04O3N2 & $1.020\,\pm\,0.004$ & $0.096\pm0.005$ & $18.09\,\pm\,0.21$ & $-0.451\pm0.003$ & ~$8.2\,\pm\,0.4$ 
& $-0.56\pm0.03$ & $-0.18\pm0.05$ \\ 
\hline
\multicolumn{11}{c}{$z>0$: $a(z)\,=\,a(0)\,(1+z)^{\gamma(a)}$, $w(z)\,=\,w(0)\,(1+z)^{\delta(w)}$} \\
 & & & & & & & & & \multicolumn{1}{c}{$\gamma(a)$} & \multicolumn{1}{c}{$\delta(w)$} \\
\multicolumn{1}{c}{(1)} &
\multicolumn{1}{c}{(2)} &
\multicolumn{1}{c}{(3)} &
\multicolumn{1}{c}{(4)} &
\multicolumn{1}{c}{(5)} &
\multicolumn{1}{c}{(6)} &
\multicolumn{1}{c}{(7)} &
\multicolumn{1}{c}{(8)} &
\multicolumn{1}{c}{(9)} &
\multicolumn{1}{c}{(10)} &
\multicolumn{1}{c}{(11)} \\
% 28/6/2016 redone 
MEGA$^{\mathrm e}$      & 25 & 0.046 & 
D02  & $0.994\,\pm\,0.001$ & $0.036\pm0.002$ & $11.25\,\pm\,0.11$ & $-0.449\pm0.002$ & ~$6.3\,\pm\,0.2$ 
& $0.46\pm0.06$ & ~$0.05\pm0.06$ \\ 
MEGA$^{\mathrm e}$      & 25 & 0.039 & 
PP04N2   & $1.136\,\pm\,0.005$ & $0.104\pm0.005$ & $20.55\,\pm\,0.26$ & $-0.413\pm0.003$ & $13.1\,\pm\,0.8$ 
& $0.68\pm0.04$ & $-0.07\pm0.04$ \\ 
MEGA$^{\mathrm e}$      & 25 & 0.049 & 
PP04O3N2 & $1.020\,\pm\,0.004$ & $0.096\pm0.005$ & $18.09\,\pm\,0.21$ & $-0.451\pm0.003$ & ~$8.2\,\pm\,0.4$ 
& $0.71\pm0.05$ & $-0.08\pm0.06$ \\ 
\\
Schechter$^{\mathrm f}$ & 28 & 0.035 & 
D02  & $0.994\,\pm\,0.001$ & $0.036\pm0.002$ & $11.25\,\pm\,0.11$ & $-0.449\pm0.002$ & ~$6.3\,\pm\,0.2$ 
& $0.48\pm0.04$ & ~$0.00\pm0.04$ \\ 
Schechter$^{\mathrm f}$ & 28 & 0.042 & 
PP04N2   & $1.136\,\pm\,0.005$ & $0.104\pm0.005$ & $20.55\,\pm\,0.26$ & $-0.413\pm0.003$ & $13.1\,\pm\,0.8$ 
& $0.73\pm0.04$ & $-0.19\pm0.04$ \\ 
Schechter$^{\mathrm f}$ & 28 & 0.049 & 
PP04O3N2 & $1.020\,\pm\,0.004$ & $0.096\pm0.005$ & $18.09\,\pm\,0.21$ & $-0.451\pm0.003$ & ~$8.2\,\pm\,0.4$ 
& $0.75\pm0.05$ & $-0.19\pm0.05$ \\ 
\\
\hline 
\multicolumn{11}{c}{$z>0$: $\mu(z)\,=\mu(0)\,[\,({\rm M}_{\rm star}/{10^{9.5}})^{\gamma(\mu)} \ (1+z)^{\delta(\mu)}$\,]$^{-1}$} \\
 & & & & & & & & & \multicolumn{1}{c}{$\gamma(\mu)$} & \multicolumn{1}{c}{$\delta(\mu)$} \\
\multicolumn{1}{c}{(1)} &
\multicolumn{1}{c}{(2)} &
\multicolumn{1}{c}{(3)} &
\multicolumn{1}{c}{(4)} &
\multicolumn{1}{c}{(5)} &
\multicolumn{1}{c}{(6)} &
\multicolumn{1}{c}{(7)} &
\multicolumn{1}{c}{(8)} &
\multicolumn{1}{c}{(9)} &
\multicolumn{1}{c}{(10)} &
\multicolumn{1}{c}{(11)} \\
% 28/6/2016, 4/7/2016 redone
MEGA$^{\mathrm e}$      & 25 & 0.062 & 
D02  & $0.994\,\pm\,0.001$ & $0.036\pm0.002$ & $11.25\,\pm\,0.11$ & $-0.449\pm0.002$ & ~$6.3\,\pm\,0.2$ 
& $-0.16\pm0.05$ & $-0.10\pm0.08$ \\ 
MEGA$^{\mathrm e}$      & 25 & 0.039 & 
PP04N2   & $1.136\,\pm\,0.005$ & $0.104\pm0.005$ & $20.55\,\pm\,0.26$ & $-0.413\pm0.003$ & $13.1\,\pm\,0.8$ 
& $-0.44\pm0.05$ & $-0.14\pm0.08$ \\ 
MEGA$^{\mathrm e}$      & 25 & 0.048 & 
PP04O3N2 & $1.020\,\pm\,0.004$ & $0.096\pm0.005$ & $18.09\,\pm\,0.21$ & $-0.451\pm0.003$ & ~$8.2\,\pm\,0.4$ 
& $-0.36\pm0.05$ & $-0.05\pm0.08$ \\ 
\\
Schechter$^{\mathrm f}$ & 28 & 0.035 & 
D02  & $0.994\,\pm\,0.001$ & $0.036\pm0.002$ & $11.25\,\pm\,0.11$ & $-0.449\pm0.002$ & ~$6.3\,\pm\,0.2$ 
& $-0.19\pm0.03$ & $-0.25\pm0.04$ \\ 
Schechter$^{\mathrm f}$ & 28 & 0.026 & 
PP04N2   & $1.136\,\pm\,0.005$ & $0.104\pm0.005$ & $20.55\,\pm\,0.26$ & $-0.413\pm0.003$ & $13.1\,\pm\,0.8$ 
& $-0.50\pm0.03$ & $-0.20\pm0.05$ \\ 
Schechter$^{\mathrm f}$ & 28 & 0.032 & 
PP04O3N2 & $1.020\,\pm\,0.004$ & $0.096\pm0.005$ & $18.09\,\pm\,0.21$ & $-0.451\pm0.003$ & ~$8.2\,\pm\,0.4$ 
& $-0.42\pm0.03$ & $-0.13\pm0.05$ \\ 
\hline
\label{tab:models} 
\end{tabular} 
}
\begin{flushleft}
$^{\mathrm a}$~The free parameters in the $z\sim0$ FPZ models are:
$a\,=\,a_{\rm coeff}\,({M_{\rm star}}/{10^{10.75}})^{a_{\rm pow}}$;
$w\,=\,w_{\rm coeff}\,({M_{\rm star}}/{10^{9.0}})^{w_{\rm pow}}$;
\epsf.
At $z>0$, there are three classes of best-fit models as described in the text: one that
depends on the variation of the inverse of the SF timescale (\epsf);
%($\epsilon_{*}(z)\,=\,({\rm M}_{\rm star}/{10^{9.5}})^\gamma \ (1+z)^\delta$), 
one that depends on the variations of the inflow ($a$) and outflow ($w$) amplitudes;
and one that depends on $\mu$, the relative gas content. \\
%($a(z)\,=\,a_{\rm coeff}\,(1+z)^\alpha$, $w(z)\,=\,w_{\rm coeff}\,(1+z)^\beta$). \\
%we have introduced a scalefactor for \epsf\ %$\epsilon_{SF}$, namely a multiplicative coefficient 
%that varies with \mstar, $({M_{\rm star}}/{10^{9.5}})^{\gamma}$,
%and one that varies with the age of the universe at redshift $z$, $\tau(z)$,
%$(1-\tau(z)/\tau(0))^\delta$, where $\tau(0)$ corresponds to the age of the universe at $z\,=\,0$. \\
$^{\mathrm b}$~Residual standard error of global fit. \\
$^{\mathrm c}$~D02 corresponds to \citet{denicolo02};
PP04N2 and PP04O3N2 to nitrogen-based and oxygen+nitrogen based calibrations by \citet{pettini04}. \\
$^{\mathrm d}$~These fits were obtained for the MEGA galaxies with $z< 0.1$.\\
$^{\mathrm e}$~We have adopted the SDSS10 $z\sim0$ best-fit model parameters at $z\sim0$ in order to fit FPZ($z>0$);
only galaxies with $z>0.1$ are included in the binned data. \\
$^{\mathrm f}$~These are fits of the model to mean \mstar\ and mean SFR using
integrals of double Schechter functions taken from \citet{ilbert13}, and assuming that sSFR varies with redshift
as the star-forming galaxies observed by \citet{karim11}; there are 28 dof in these fits
(6 redshift bins $\times$ 5 mass bins, since we do not use the lowest-mass bin, and two of the mass bins are empty at the given redshift bin).
The redshifts bins and the metallicity behavior with redshift to emulate are taken from mean trends of the MEGA dataset.
We have adopted the SDSS10 $z\sim0$ best-fit model parameters at $z\sim0$ in order to fit FPZ($z>0$).
See text for more details. \\
\end{flushleft}
\end{center}
\end{table*}
%together with the best-fit theoretical inflow ($a$, column 2),
%outflow ($w$, column 3), and star-formation timescale (\epsf, column 4)
%FMR parameters. 

%Despite the different O/H calibrations,
Despite the differences (O/H calibrations, stellar yields, IMF),
our new best-fit FPZ model parameters are generally similar to those of \citet{dayal13}.
The small \mstar\ dependence of the accretion term, and its small amplitude
are consistent with previous values;
for the galactic winds, however,
we find however a slightly larger coefficient and a slightly steeper
\mstar\ dependence ($\sim -0.4$ vs. $\sim -0.33$).
The inverse SF timescales \epsfinv\ are also smaller than in \citet{dayal13},
corresponding to $\sim 100$\,Myr vs. $\sim 600$\,Myr previously. 
Part of the differences in the best-fit parameters may be due to 
the choice of the model IMF and our assumptions
for the O/H calibration; here we use the \citet{asplund09} values of
\logoh\,=\,8.69 for solar oxygen abundance, and a metal mass fraction Z/H of 0.0198\footnote{This corresponds
to an oxygen abundance of $4.9\times10^{-4}$ while
\citet{dayal13} used $7.9\times10^{-4}$.}. 
Nevertheless, despite this and the difference in the fitting methods, the results
are in reasonably good agreement.

The mass loading factor for galactic winds, $w$, found by the SDSS10 best-fit model
is roughly consistent with those found by the cosmological hydrodynamic
simulations of \citet{dave11}, despite their different underlying assumptions
about the relation of SFR, gas accretion, and galactic winds. 
They assume an ``equilibrium'' model (further described below, Sect. \ref{sec:comparison})
in which the infall rate $a$ is balanced by the sum of the SFR
$\psi$ and the outflow $w$.
The results of their simulations show a trend with stellar mass of $w$ $\propto {\rm M}_{\rm star}^{-0.33}$,
or momentum-driven winds, consistent with the MEGA best fit, and
only slightly flatter than the power-law index $a_{\rm pow}\sim-0.4$
given by the SDSS10 best fit.
This power-law \mstar\ dependence of $\sim -0.4$ corresponds roughly to the low-mass linear portion of SDSS10 MZR 
with a slope of $\sim 0.38$ (Paper I).
%slope $\sim 0.38$ shown in Fig. \ref{fig:mzrsdss}; because of the stronger \mstar\ dependence
%and the higher $w_{\rm coeff}$ for winds relative to accretion
%this low-mass regime is where galactic winds are dominating the shape of the MZR.

In contrast, the accretion or infall parameter $a$ is significantly
different from what is found in some equilibrium models.
While the infall amplitude $a$ is required to be larger than the outflow $w$
in such models, $a\,=\,(w+1)\,\psi$ \citep[see, e.g.,][]{dave11}, 
we find a much smaller $a$ value, roughly a factor of 10 smaller than
$w$, and with a much shallower \mstar\ dependence.
These differences will be discussed further in Sect. \ref{sec:comparison}.
In any case,
our model is able to reproduce O/H of the $\sim$80\,000 SDSS10 galaxies
to within $\sim 0.05-0.07$\,dex, lending credibility to our approach.

The quantity \epsfinv\ corresponds approximately to a gas depletion time,
with some caveats.
Observationally, gas depletion times, \taud, are inferred from 
the measured ratio of SFR and gas mass, \taud\,=\,$\psi$/\mg, either locally as surface densities,
or globally integrated over the entire galaxy \citep[e.g.,][]{bigiel11,saintonge11,huang14,genzel15,hunt15}.
For spiral galaxies, typical depletion times \taud\ for the \htwo\ component
are $\sim$2.4\,Gyr \citep[e.g.,][]{bigiel11} but
can be as small as 50\,Myr in metal-poor starbursts \citep{hunt15} or as large as $\sim$10\,Gyr
in more quiescent systems \citep[e.g.,][]{saintonge11,bothwell14}. 
On the other hand, atomic gas \hi\ depletion times are relatively constant,
$\sim$3.4\,Gyr \citep{schiminovich10}.
The values we find for \epsfinv\ ($\sim 100-200$\,Myr for SDSS10, $\sim 400-700$\,Myr for MEGA) are shorter than either of these.
%(\epsfinv\ $\sim$500\,Myr for the MEGA sample, see below).

The reason for this disagreement can be understood as follows:
the constant of proportionality \epsf\ relating gas mass and SFR
(see Eqn. (\ref{eqn:sfr}))
is in reality the ratio of a (dimensionless) star-formation efficiency
\epff\ and a SF timescale, \tff\
\citep[e.g.,][]{krumholz09,krumholz12}, where the subscript ``ff'' refers to a free-fall
or dynamical timescale.
Physically, \epff\ gives the fraction of gas converted into stars over
a dynamical timescale, in a process
that is usually highly inefficient with \epff$\sim 0.01-0.05$
\citep[e.g.,][]{krumholz07}.
With such efficiencies and assuming that \epsf\,=\,\epff/\tff\,
the SDSS10 fitted values of \epsfinv\ would give \tff$\sim 1-8$\,Myr.
These times are unrealistically small for typical giant molecular clouds or spiral
disks, which instead have typical dynamical times \tff$\sim$15-20\,Myr \citep{krumholz12}.

However, what we call \mg\ is in fact only the gas that enriches the ISM with metals
as it is recycled through star formation; it is not the total gas available that would
be observed.
In other words, \mg(model)\,=\,\mg(observed)\,\epff, or from Eqn. (\ref{eqn:sfr}):
\begin{equation}
\epsilon_{*}\,=\,\frac{\psi}{M_g}\,=\frac{\psi}{M_g{\rm (observed)}\,\epsilon_{\rm ff}}\,=\,\frac{1}{\tau_{\rm depl}\,\epsilon_{\rm ff}}.
\label{eqn:epsilon}
\end{equation}
\noindent
We can use this equation to calculate typical gas depletion times as estimated from
our fits of the SDSS10 galaxies.
Assuming \epff$\sim$0.03,
we find \tff$\sim$3\,Gyr, similar to 
typical gas depletion times for both the molecular and the atomic component\footnote{To explore
the molecular gas fraction necessary to form stars would be an important addition
to this discussion, but is beyond the scope of this work.}.

Comparison of the MEGA results with those of SDSS10
shows that the amplitude and power-law index for the wind parameters are consistent with SDSS10,
but the accretion parameters and the inverse SF timescales, \epsf, differ.
As can be seen in Fig. \ref{fig:fmrmega}, the high-mass end of the SFR dependency of O/H
falls off more steeply for the MEGA galaxies than for SDSS10.
For the PP04N2 and PP04O3N2 calibrations,
gas accretion for the MEGA galaxies seems to increase with decreasing \mstar,
in contrast to the behavior of the SDSS10 sample;
because of the highly star-forming nature of the MEGA dataset, this could be
telling us that the lower-mass galaxies are more gas rich than higher-mass ones. 
On the other hand, the SF timescales implied by the FPZ model fit to the MEGA
galaxies are longer than for SDSS10, although with relatively large uncertainties.
More generally,
the SDSS10 best-fit parameters are in all cases much better determined than those 
for the MEGA dataset, probably because of the much larger number of galaxies in the former.
Therefore, in what follows, we will adopt the best-fit FPZ parameters for the SDSS10 $z\,\simeq\,0$
galaxies to extend the model to high redshift.
Table \ref{tab:models} gives the best-fit FPZ model parameters for the SDSS10
and MEGA datasets. 

%%%%%%%%%%%%%%%%%%%%%%%%
%\begin{figure*}[!htbp]
\begin{figure*}
\vspace{\baselineskip}
\hbox{
\includegraphics[height=0.45\textwidth]{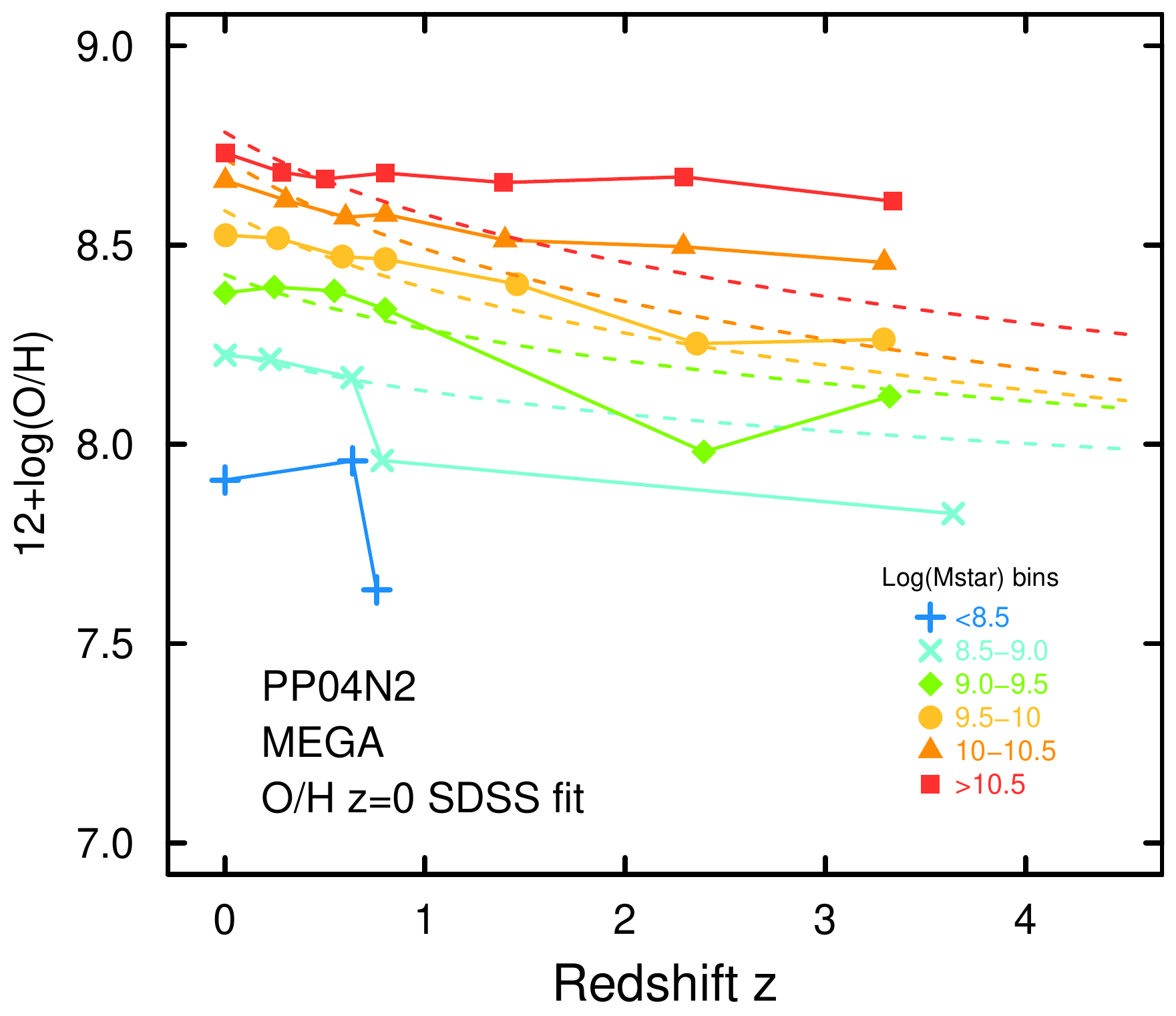}
%\hspace{0.3cm}
 \includegraphics[height=0.45\textwidth]{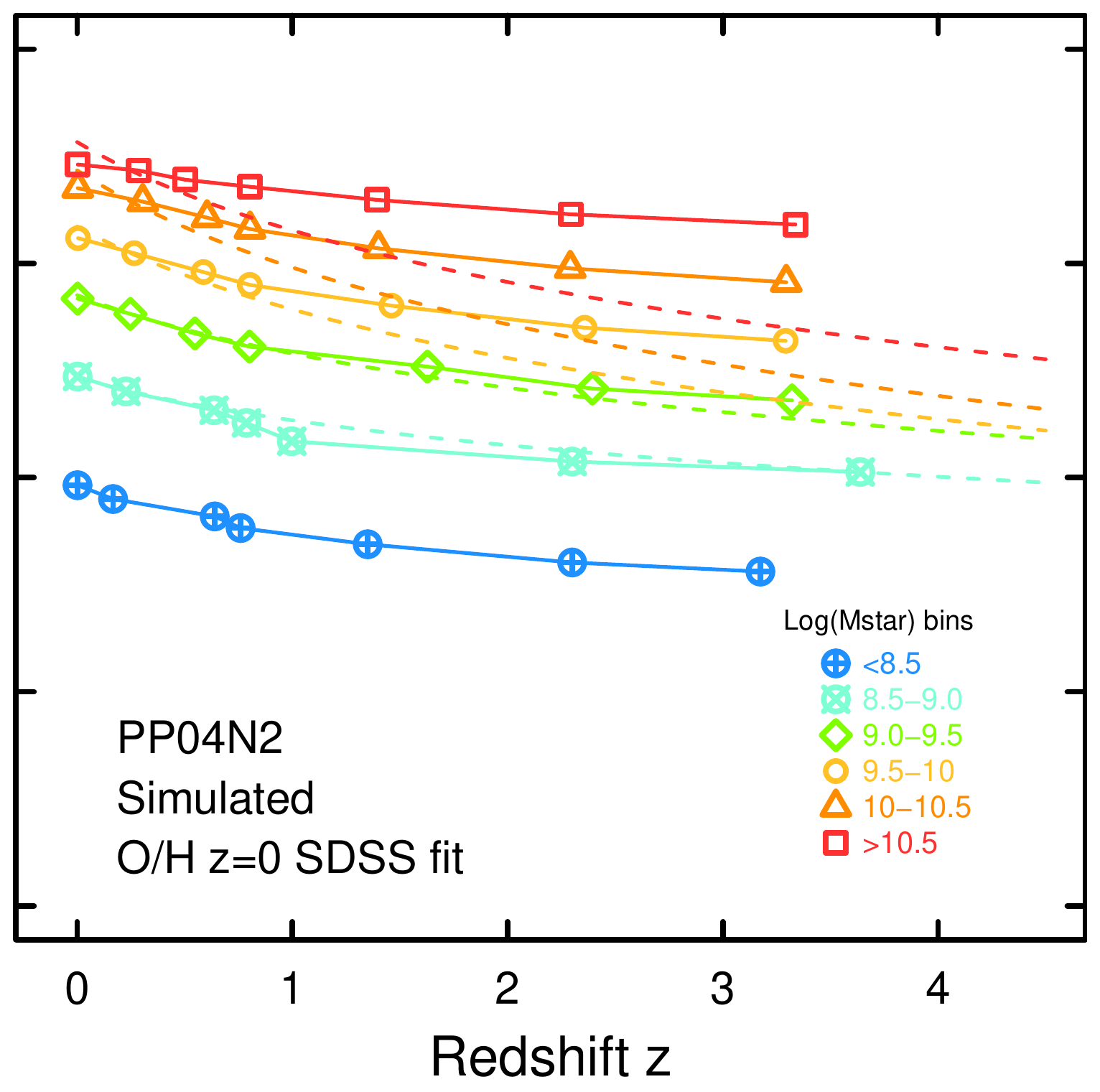}
}
\caption{Binned measurements of \logoh\ estimated from SFR and the FPZ model
at $z\,\simeq\,0$ as a function of redshift
for the MEGA dataset (left panel) and the simulated dataset (right). 
In both panels, the symbols correspond to the values of \logoh\ that would be
inferred using the FPZ($z\simeq 0$) model applied to the median SFR and \mstar\
within each redshift bin (according to binned \mstar\ values);
the curves show the average behavior of the MEGA dataset as reported in Paper I.
%Fig. \ref{fig:ohssfrvsz}.
See text for a description of the simulated galaxy populations shown in the right panel.
}
\label{fig:nofmrvsz}
\end{figure*}

\subsection{Comparison with previous work at $z\simeq 0$}
\label{sec:comparison}

As discussed in Paper I, the intention of the FPZ is similar to that of the
``Fundamental Metallicity Relation'' \citep[``FMR'', ][]{mannucci10},
namely to relate the metal content of galaxies to their star-formation activity.
Nevertheless, the two formulations are different;
the planar formulation of the FPZ, accurate to $\sim$0.16\,dex, extends 
to galaxies with \mstar\ as low as $\sim 10^6$\,\msun\ and to redshifts
$\la 3.7$ (see Sect. \ref{sec:coevo} and Paper I). 
In contrast,
the extension to lower stellar masses of the FMR 
is quadratic \citep{mannucci11}, and, as shown in Paper I, has a larger 
mean dispersion than the FPZ and a significant offset for the MEGA sample
which spans a much broader parameter space than the original SDSS dataset 
at $z\sim 0$ for which the FMR was developed.

Due to its success, which by including SFR reduced significantly
the dispersion for \logoh\ of the original dataset relative to only stellar mass,
many models have focused on reproducing the FMR at $z\simeq 0$ by
\citet{mannucci10}.
%with varying degrees of success.
Because of the underlying notion that metallicity is somehow connected to SFR,
common to both the FPZ and the FMR formulations,
before pursuing our model for the FPZ,
here we discuss previous attempts to theoretically understand the FMR.

%Various degrees of success of reproducing the 
%FPZ at $z\simeq 0$ have been achieved so far
%with other models.
Despite using simulation runs calibrated to a number $z \simeq 0$
observables (stellar mass function, stellar mass-size relation and the stellar
mass-black hole relation) the current gold-standard EAGLE simulations are
unable to reproduce the local FMR; whilst reproducing observations to within
0.15\,dex for high mass (\mstar$\la 10^{10}$\,\msun) galaxies, they severely
over-estimate the metal content of lower mass halos by as much as 0.4\,dex
\citep{lagos16}. 
Given the sub-grid limitations of simulation, most effort
has thus been diverted to developing an analytic understanding of the $z \simeq 0$
FMR. 

We briefly discuss the two previous works most close in spirit to ours. The
\citet{dave12} model assumes every galaxy to be in ``equilibrium" with the
infall rate exactly balancing the gas lost in outflows and star formation.
While this model correctly yields a metallicity that is independent of star
formation for high-mass galaxies, it fails to capture the metallicity downturn
observed at low mass. Interestingly, however, this model finds
outflows to be momentum driven with a mass power-law dependence of $\sim -0.33$, %consistent with our findings.
which however we obtain only for the MEGA sample; 
the SDSS10 slopes are slightly steeper than this, $\sim -0.4$.
\citet{peeples11} find that reproducing the local mass-metallicity relation
requires a high metal-expulsion efficiency (compared to expelled gas,
``preferentially metal-enriched winds") that scales steeply with increasing \mstar.
Although this might yield the observed FMR (or FPZ) for low-mass galaxies, it is doubtful
if such an assumption could reproduce the constant metallicity observed for the
most massive systems, independently of SFRs ranging over five
orders of magnitude presented in the MEGA sample. 

\citet{lilly13} use the
``gas-regulator" model wherein stars are embedded in a gas reservoir
whose mass is increased by infall, and the SFR being proportional to the gas
mass available at any time. Assuming both the star formation efficiency and
outflow rate to be independent of the halo mass (the ``ideal regulator"), these
authors find the metallicity to be largely independent of the
evolutionary path because of gas continually ``flushing" the system. This is quite
contrary to our model that requires outflows to scale with the halo mass, and
finds the metallicity to explicitly depend on the gas evolution history.

%%%%%%%%%%%%%%%%%%%%%%%%
%\begin{figure*}[!htbp]
\begin{figure*}
\vspace{\baselineskip}
\hbox{
\includegraphics[height=0.45\textwidth]{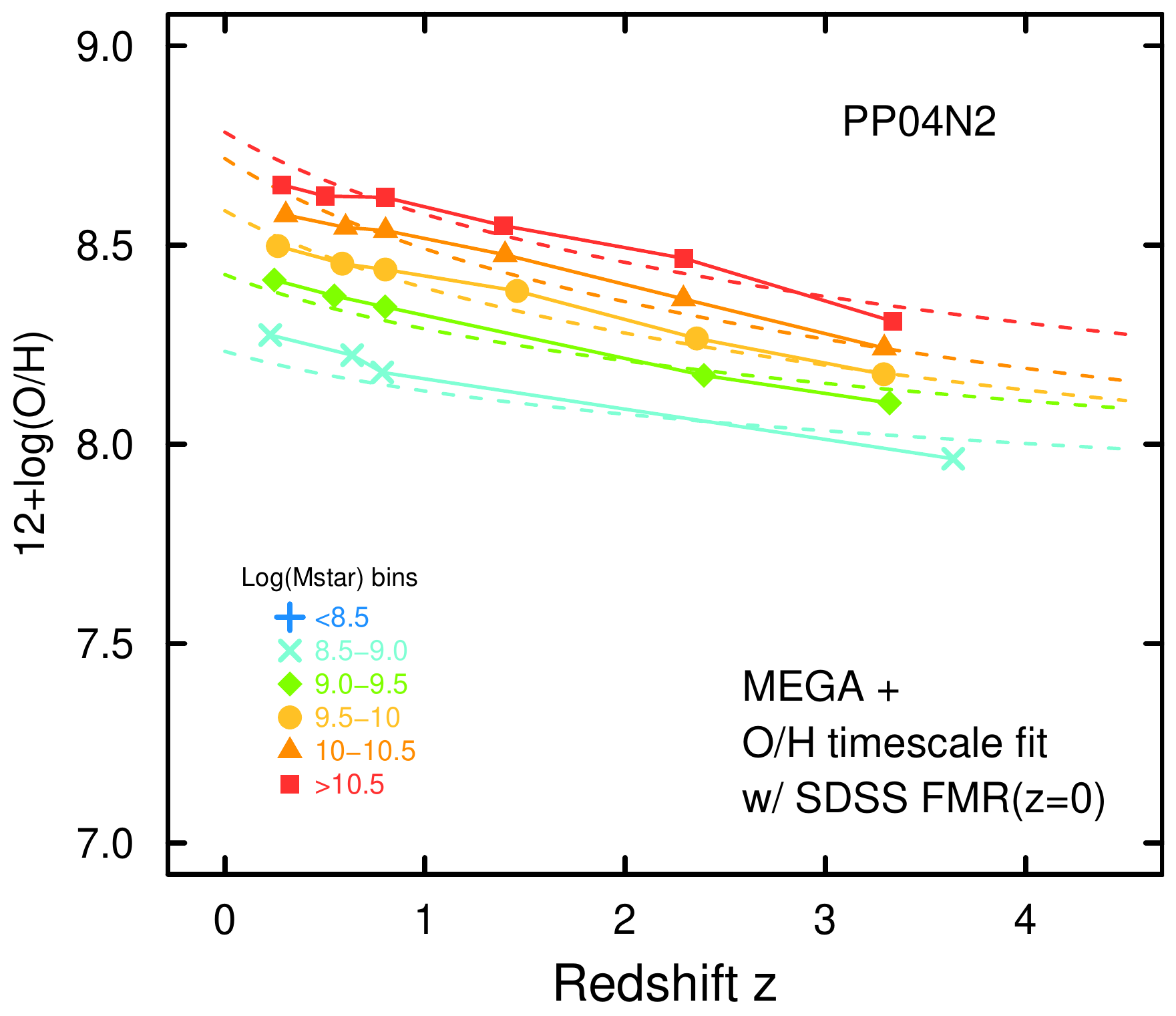}
%\hspace{0.3cm}
 \includegraphics[height=0.45\textwidth]{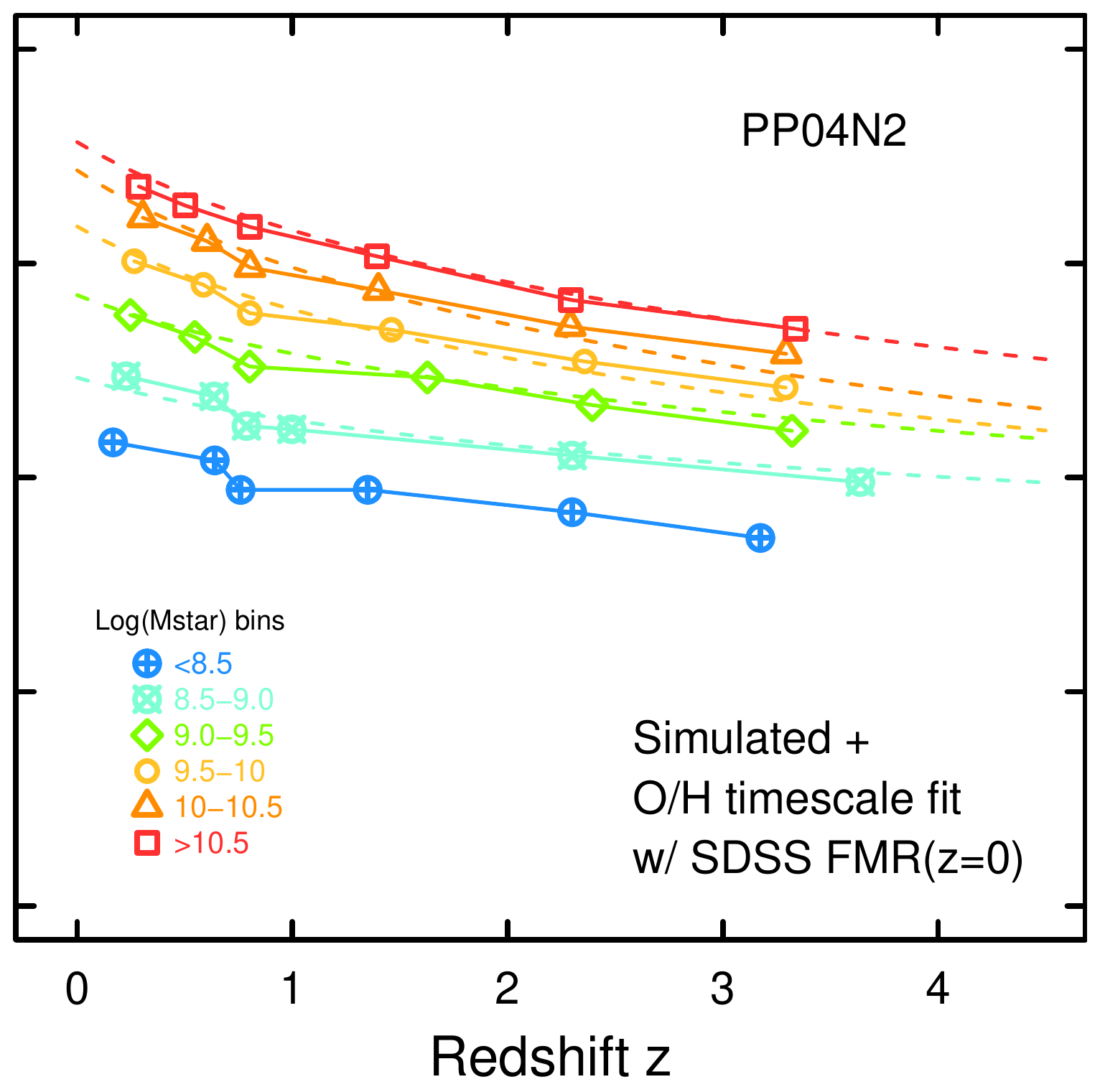}
}
\caption{Binned measurements of \logoh\ estimated from SFR and the FPZ($z$) model
with \epsf($z$) added to the FMZ($z=0$) model as a function of redshift
for the MEGA dataset (left panel) and the simulated dataset (right). 
In both panels, the symbols correspond to the (PP04N2) values of \logoh\ that would be
inferred using the FPZ($z\simeq 0$) model applied to the median SFR and \mstar\
within each redshift bin (according to binned \mstar\ values);
the curves show the average behavior of the MEGA dataset as reported in Paper I.
%Fig. \ref{fig:ohssfrvsz}.
See text for a description of the simulated galaxy populations shown in the right panel.
}
\label{fig:timescalefmrvsz}
\end{figure*}

%%%%%%%%%%%%%%%%%%%%%%%%
%\begin{figure*}[!htbp]
\begin{figure*}
\vspace{\baselineskip}
\hbox{
\includegraphics[height=0.45\textwidth]{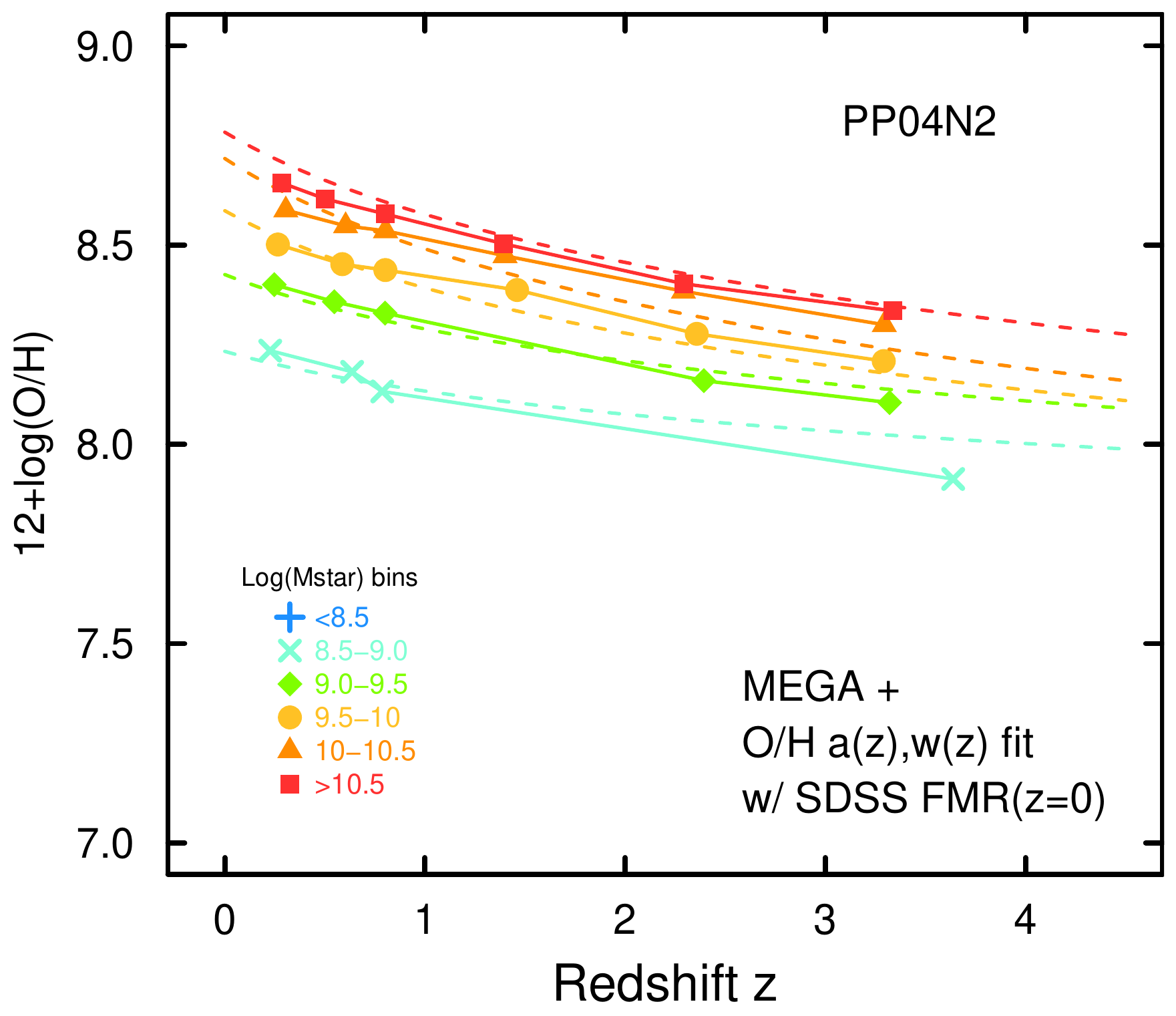}
%\hspace{0.3cm}
 \includegraphics[height=0.45\textwidth]{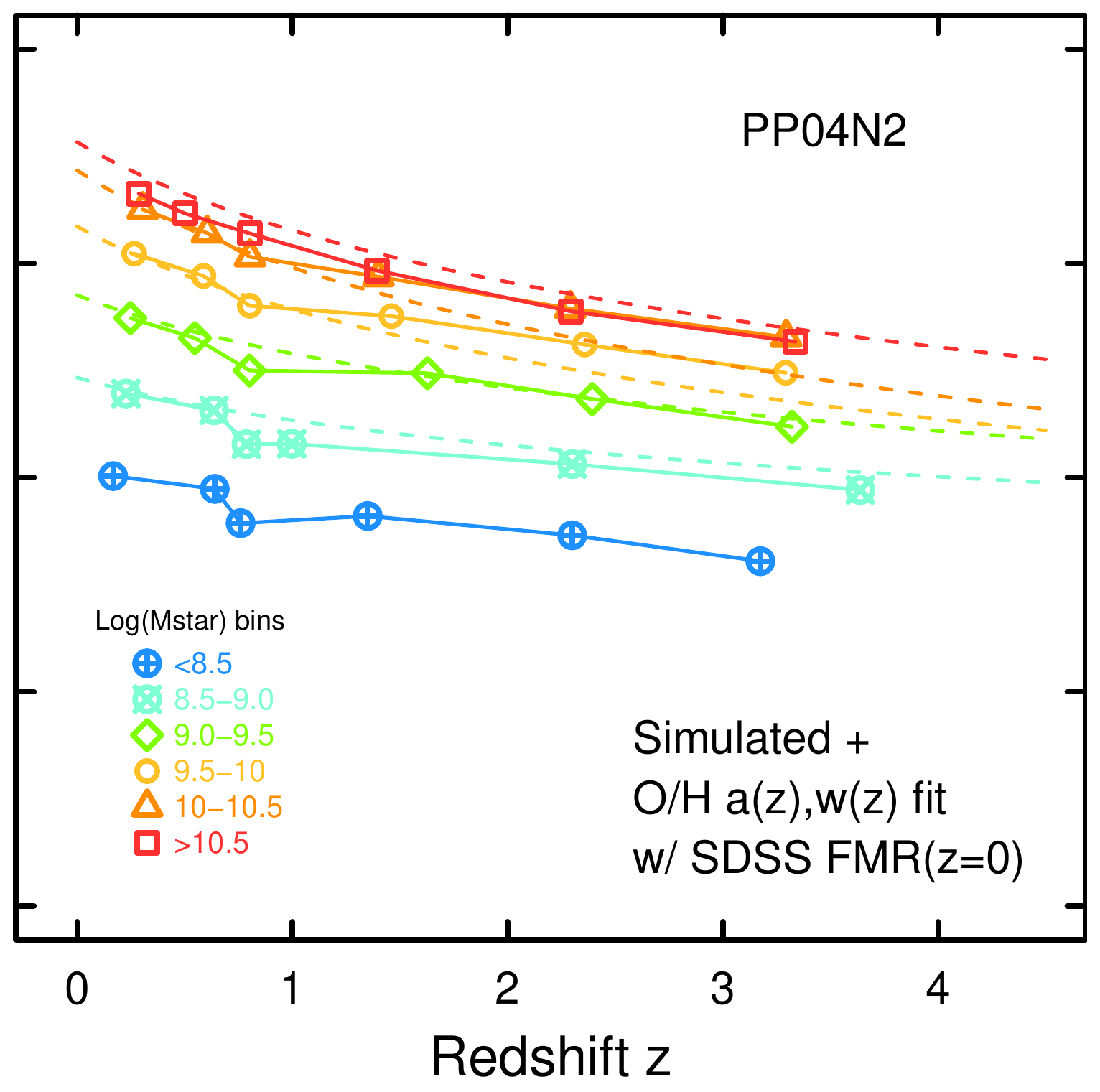}
}
\caption{Binned measurements of \logoh\ estimated from SFR and the FPZ($z$) model
with $a(z)$, $w(z)$ added to the FMZ($z=0$) model as a function of redshift
for the MEGA dataset (left panel) and the simulated dataset (right). 
In both panels, the symbols correspond to the (PP04N2) values of \logoh\ that would be
inferred using the FPZ($z\simeq 0$) model applied to the median SFR and \mstar\
within each redshift bin (according to binned \mstar\ values);
the curves show the average behavior of the MEGA dataset as reported in Paper I.
%Fig. \ref{fig:ohssfrvsz}.
See text for a description of the simulated galaxy populations shown in the right panel.
}
\label{fig:coefffmrvsz}
%\end{figure*}
%%%%%%%%%%%%%%%%%%%%%%%%
%\begin{figure*}[!htbp]
%\begin{figure*}
\setcounter{figure}{5}
\vspace{\baselineskip}
\hbox{
\includegraphics[height=0.45\textwidth]{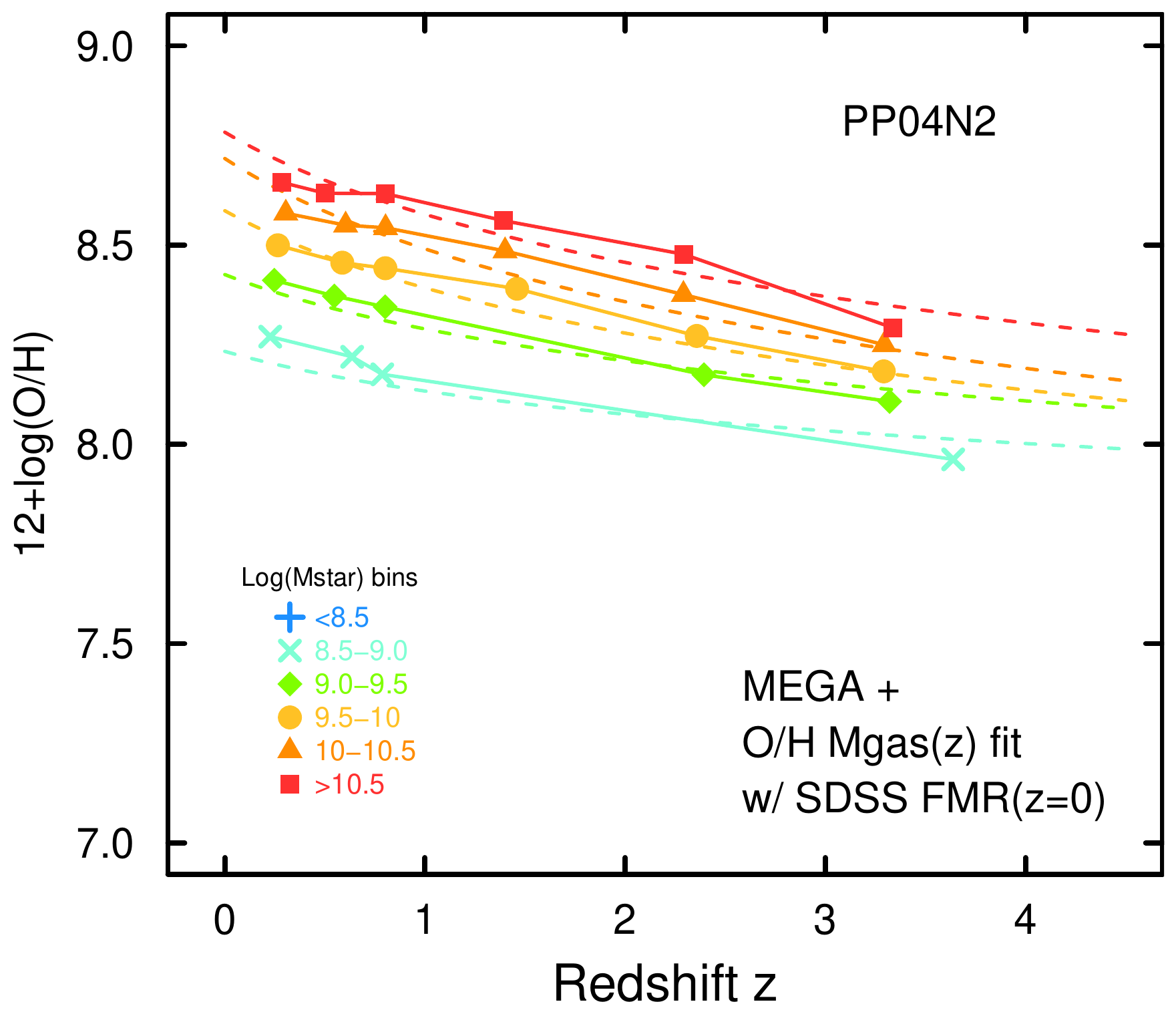}
%\hspace{0.3cm}
 \includegraphics[height=0.45\textwidth]{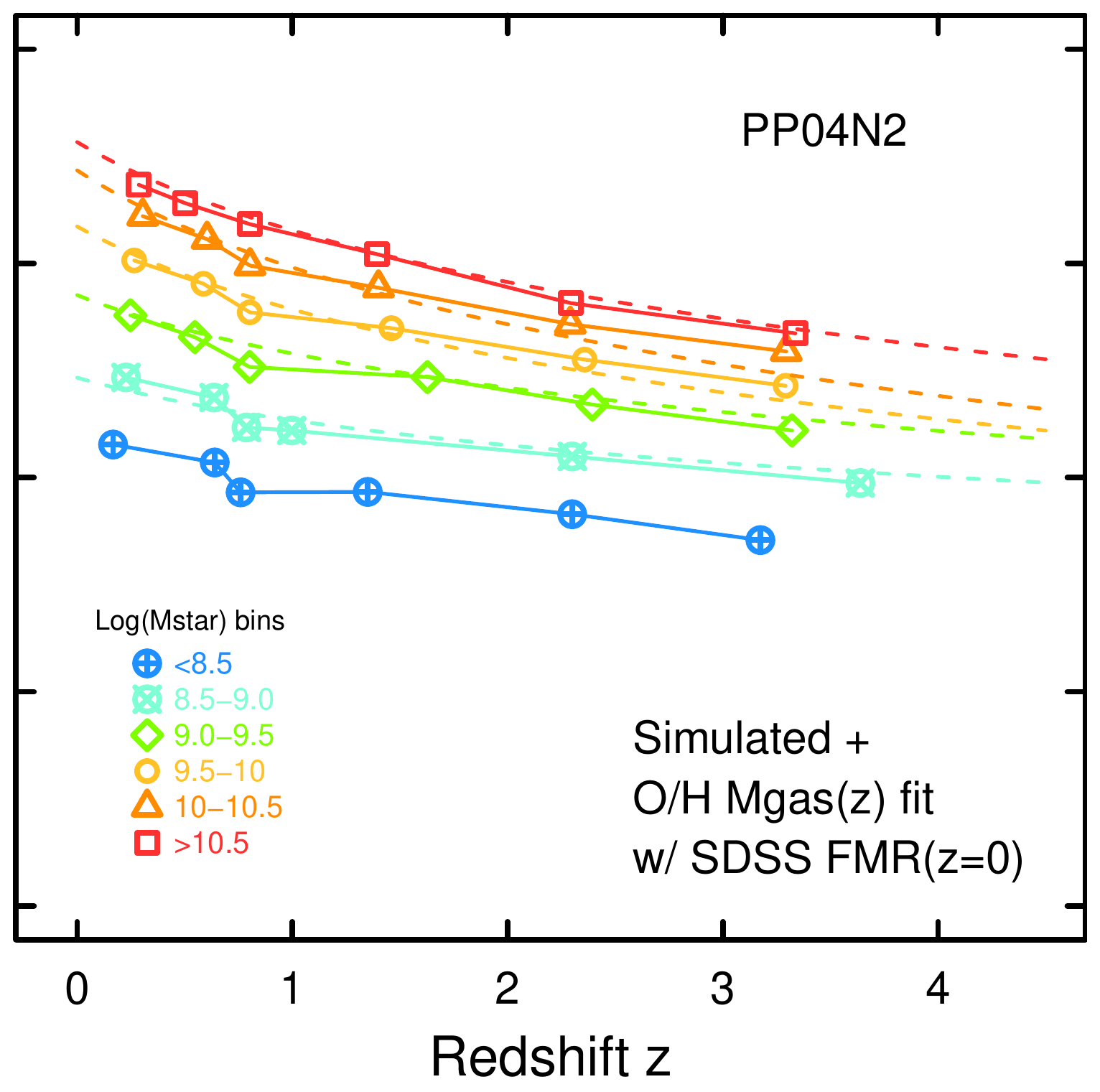}
}
\caption{Binned measurements of \logoh\ estimated from SFR and the FPZ($z$) model
with \mg($z$) added to the FMZ($z=0$) model as a function of redshift
for the MEGA dataset (left panel) and the simulated dataset (right). 
In both panels, the symbols correspond to the (PP04N2) values of \logoh\ that would be
inferred using the FPZ($z\simeq 0$) model applied to the median SFR and \mstar\
within each redshift bin (according to binned \mstar\ values);
the curves show the average behavior of the MEGA dataset as reported in Paper I.
%Fig. \ref{fig:ohssfrvsz}.
See text for a description of the simulated galaxy populations shown in the right panel.
}
\label{fig:mgasfmrvsz}
\end{figure*}

\subsection{Testing the $z \simeq 0$ model at higher redshift}
\label{sec:testfmrhighz}

In order to extend the FPZ model to $z>0$, we first test whether the SDSS10 best-fit
$z\simeq 0$ parameters apply to the data at higher redshift.
Such behavior would be expected if the lower metallicities at high redshift
were compensated for merely by the more intense star-formation activity, which could
%power higher levels of accretion of pristine gas, 
power larger outflow of metals through stronger galactic winds. 
The left panel of Fig. \ref{fig:nofmrvsz} shows the behavior of the MEGA sample if we 
infer metallicity through the FPZ($z\simeq 0$) model parameters. 
The curves are the fits to the observed O/H of the
MEGA dataset as a function of redshift as reported in Paper I; %Fig. \ref{fig:ohssfrvsz}
%(see also Table \ref{tab:ohvszbins});
the symbols correspond to the values of \logoh\ that would be inferred from the FPZ
$z\,=\,0$ model applied to the median SFR and \mstar\ in each redshift bin.
In all \mstar\ bins, up to $z \sim 0.7$ the FPZ correctly predicts
the observed O/H.
However, for higher redshifts, given the observed (median) SFRs,
the FPZ model overpredicts the observed O/H, 
in particular for the highest \mstar\ mass bins where the
discrepancy is $\ga$0.4\,dex.
On the other hand, in the lowest mass bin (log(\mstar/\msun)\,=8.5-9.0) 
O/H is underestimated by $\sim$0.1\,dex.
Similar tensions relative to the $z=0$ model by \citet{dayal13}
are seen also in the sample of $z\sim 2$ galaxies by 
\citet{grasshorn16} where they find that galaxies with \mstar$\ga 10^9$\,\msun\
have higher O/H relative to their local counterparts, while lower-mass systems
have lower abundances.

Because the high SFRs of the MEGA dataset may not be fully representative of typical high-$z$ 
``main-sequence'' galaxy populations (see Paper I), %Sect. \ref{sec:massvariation}), 
as a check, we have
simulated truly main-sequence galaxy populations as a function of redshift using the Galaxy Stellar Mass Functions (GSMFs) by
\citet{ilbert13} for $z>0$ and for $z \simeq 0$ the GSMF from \citet{baldry12}.
For a given \mstar, the SFR has been inferred
by using the SFMS formulation of \citet{speagle14} as a function of cosmic time.  % (see Fig. \ref{fig:ms}).
For each redshift bin, integration over the appropriate weighted GSMFs
was performed to obtain the mean \mstar\ and the mean SFR at that redshift for
main-sequence populations.
Using these values of \mstar\ and SFR,
we then calculated O/H using the FPZ($z\simeq 0$) model parameters as for the MEGA data.
The result is shown in the right panel of Fig. \ref{fig:nofmrvsz}, where
the metallicities predicted from the FPZ($z\simeq 0$) model are compared to 
%the metallicities expected for a given \mstar\ and $z$ judging from the best-fit
the mean observed O/H behavior (shown by curves) of the MEGA dataset as reported in Paper I. % as given in the left panel of Fig. \ref{fig:ohssfrvsz}.
As in the left panel, symbols correspond to the metallicities that would be
inferred for the simulated galaxies of a given \mstar\ and SFR(\mstar,$z$) from the FPZ($z\simeq 0$) model within each \mstar\ and redshift bin. 
Despite the two very different approaches, both the simulated galaxies (right panel
of Fig. \ref{fig:nofmrvsz}) and the MEGA galaxies (left panel) show similar
behavior.
The FPZ($z\simeq 0$) parameters are successful to $z \sim 0.7$, but fail in 
the same way as for the MEGA dataset at higher $z$.
The consistency of the simulations and the MEGA galaxies is encouraging,
and in the next section we use both methods to extend the FPZ model to high redshift.

\section{Extending the model to higher redshift}
\label{sec:fmrhighz}

As illustrated in the previous section, the higher SFRs at $z>0$ are insufficient
by themselves through the FPZ($z\simeq 0$) model to lower the metallicities to the
levels observed.
We have thus investigated three avenues of adapting the FPZ model to $z>0$:
(a) changes in \epsf, since timescales and/or star-formation efficiencies might be expected to change with redshift
(Sect. \ref{sec:epsilon_z});
(b) redshift variations of accretion and wind parameters ($a$ and $w$)
(Sect. \ref{sec:aw_z});
and (c) higher gas fractions through possible changes in the $\mu$ parameter (
Sect. \ref{sec:mgas_z}, see Eqn. (\ref{eqn:X})).
Since the discrepancies in Fig. \ref{fig:nofmrvsz} seem to depend on \mstar,
we considered separate dependencies on \mstar\ and redshift (or cosmic time).
Thus, for each approach to establish the $z$-dependent FPZ model, we fix FPZ($z\simeq 0$) to the SDSS10 parameters and introduce
a scaling factor for either \epsf, ($a$, $w$), or $\mu$. 
Thus only two parameters are to be fit: one for a possible \mstar\ scaling
for $z>0$ (\epsf, $\mu$) and one for a scaling with redshift (\epsf, $\mu$, $a$, $w$).
Although these are only a few of the many possible adaptations of the FPZ model to
high $z$, we have limited the possibilities for simplicity.

To constrain the model, we have assumed that the true metallicities of high-$z$
galaxy populations
are approximated by the mean behavior of the MEGA dataset as reported in Paper I.
The best-fitting model parameters for $z>0$ are obtained by minimizing the residuals of the %``true'' 
observed mean behavior of \logoh\ relative to the model over the various
mass and redshift bins, not including $z\simeq 0$; % for either the MEGA dataset or the simulated galaxies.
because of small numbers at high redshift,
we also do not consider the lowest mass bin (log(\mstar/\msun)$<$8.5).
The resulting degrees of freedom in the fits are 
25 for the MEGA dataset (since some \mstar\ bins are missing at $z>0$), and 
28 for the simulated galaxies (5 \mstar\ bins over 6 redshift bins, 2 free parameters). 
As mentioned above, in all high-$z$ FPZ models, we adopted the $z\simeq 0$ SDSS10 parameters
and adjusted \epsf\ (first approach), $a$ and $w$, the accretion and wind coefficients
(second), or $\mu$ (third).
Table \ref{tab:models} gives the best-fit FPZ($z$) (D02, PP04N2, PP04O3N2)
model parameters for the three
approaches which are described in the following sections.

\subsection{Redshift variation of star-formation timescales and/or efficiencies}
\label{sec:epsilon_z}

We first investigated whether scaling \epsf\ with \mstar\ and $z$ could have
the desired effect of lowering O/H at high redshift.
Several methods were explored
for introducing a scaling factor \esca\ applied as a multiplicative
factor to \epsf\ in the FPZ models.
%These include: \esca\,=
%\begin{enumerate}[(a)]
%\item
%dex(\mstar-9.5)$^\gamma$\ $(1-{\rm age}_z/{\rm age}_H)^\delta$
%\item
%dex(\mstar-9.5)$^\gamma$\ $(1+z)^\delta$
%\item
%dex($\psi$-1)$^\gamma$\ $(1-{\rm age}_z/{\rm age}_H)^\delta$
%\item
%dex($\psi$-1)$^\gamma$\ $(1+z)^\delta$
%\item
%dex(\mstar-9.5)$^\gamma$\ $({\rm age}_z/{\rm age}_H)^\delta$
%\item
%dex(\mstar-9.5)$^\gamma$\ $({\rm age}_z)^\delta$
%\item
%dex(\mstar-9.5)$^\gamma$\ $({\rm age}_z)\,\delta$
%\end{enumerate}
%In the above, ${\rm age}_z$ is the age of the Universe at redshift $z$,
%${\rm age}_H$ is the Hubble time,
%and the scaling factors of dex(9.5) for \mstar\ and dex(1) for $\psi$ (SFR) are included
%for convenience of calculation;
%$\gamma$ and $\delta$ are the parameters to be fit.
%Options (c) and (d) are a sort of mass-quenching scenario as formulated
%by \citet{peng10}.
%Only two, options (b) and (e) 
Although we studied numerous ways to scale \epsf, including
a mass-quenching scenario as formulated by \citet{peng10}, 
only two gave reasonably low mean O/H residuals, $\sim 0.03-0.04$\,dex for
both MEGA and simulated galaxies.
These correspond to \epsf($z$)\,=\,\esca\,\epsf(0) where \esca\ is given by:
\begin{enumerate}[(a)]
\item
(\mstar/$10^{9.5})^{\gamma(\epsilon_*)}$\ $(1+z)^{\delta(\epsilon_*)}$
\item
(\mstar/$10^{9.5})^{\gamma(\epsilon_*)}$\ $({\rm age}_z/{\rm age}_H)^{\delta(\epsilon_*)}$
\end{enumerate}
where \epsf(0) gives the value of \epsf\ at $z\,=\,0$,
age$_z$ and age$_H$ are the ages of the universe at redshift $z$ and $z=0$, respectively,
and $\gamma(\epsilon_*)$ and $\delta(\epsilon_*)$ are the parameters to be fit.
Since the two fits are equivalent in terms of quality (low residuals), 
we chose (a) since it is generally easier to formulate numerically.

The values of \logoh\ predicted by this model as a function of redshift
are shown in Fig. \ref{fig:timescalefmrvsz}; 
as in Fig. \ref{fig:nofmrvsz} the left panel gives the MEGA dataset
and the right panel the simulated galaxies. 
The fit is excellent for the simulated galaxies, with a mean residual $\sigma\,\sim\,0.025$\,dex
(except for D02),
but for the MEGA dataset is slightly worse, $\sigma\,=\,0.04 - 0.06$\,dex.
There are some problems in the intermediate-$z$ regime where the metallicities
inferred from the FPZ model by varying \epsf\ with mass and $z$ tend to be overestimated.
The predictions for low-mass galaxies also tend to be overestimated at low redshift.

\subsection{Redshift variation of gas accretion and galactic winds}
\label{sec:aw_z}

The second avenue of investigation relied on simultaneously scaling the gas accretion coefficient $a$
and the galactic wind coefficient $w$, in order
to quantify whether the scaling with SFR changes with redshift.
For simplicity, we used one coefficient $\gamma(a)$ to define the redshift variation of $a$,  
and another $\delta(w)$ for that of $w$:
\begin{eqnarray}
a(z)&=&a(0)\,(1+z)^{\gamma(a)} \nonumber \\
w(z)&=&w(0)\,(1+z)^{\delta(w)} \nonumber 
\end{eqnarray}
where $a(0)$ and $w(0)$ are the values of $a_{\rm coeff}$ and $w_{\rm coeff}$ at $z\,=\,0$.
The implicit assumption is that the \mstar\ dependence of $a(0)$ and $w(0)$ given by
the SDSS10 fit in Table \ref{tab:models} holds also at $z>0$.

Figure \ref{fig:coefffmrvsz} gives the results of this fit, where, as before,
the left panel shows the MEGA dataset and the right the simulated galaxies.
%The fit for the MEGA dataset is excellent, the best of all the formulations we
%tried, with a mean residual between data and predicted O/H of $\sigma\,=\,0.03$\,dex.
The fit for the MEGA dataset is good, 
with a mean residual between data and predicted O/H of $\sigma\,=\,0.04 - 0.05$\,dex, 
similar to the previous \epsf\ approach described above.
For the simulated galaxies, the fit is still good ($\sigma\,=\,0.04 - 0.05$\,dex),
although slightly worse than the \epsf\ model (except for D02 where it is better). 

\subsection{Redshift variation of gas fraction}
\label{sec:mgas_z}

The last successful formulation involved scaling the model parameter $\mu$
which basically defines the gas-mass fraction relative to its initial value
in the galaxy's evolution (see Eqns. (\ref{eqn:X}) and (\ref{eqn:mgas})).
For computational reasons, the scaling factor was introduced in the denominator
of $\mu$: 
\begin{equation}
\mu(z)\,=\,\mu(0)\ [\,({\rm M}_{\rm star}/10^{9.5})^{\gamma(\mu)}\,(1+z)^{\delta(\mu)})\,]^{-1} \nonumber
\end{equation}

The FPZ($z$) model predictions of O/H with this approach are shown in
Fig. \ref{fig:mgasfmrvsz} where the MEGA dataset is shown in the left panel
and the simulated galaxies in the right.
Again, the fit for the MEGA dataset is good, 
with mean residuals  $\sigma\,=\,0.04 - 0.06$\,dex, 
consistent with the previous two formulations.
The behavior of the MEGA galaxies is similar to the \epsf\ approach
described in Sect. \ref{sec:epsilon_z} where O/H is over-predicted by
the model in the intermediate-$z$ redshift range, although the discrepancies
at the low-mass end are slightly less pronounced.
On the other hand, this formulation of the FPZ($z$) model for the
simulated populations is excellent, with a mean residual of $\sigma\,\sim\,0.03$\,dex. 

\section{Discussion}
\label{sec:discussion}

This is one of the first works that aims to study the physics
governing the redshift-evolution of the FPZ (or FMR).
\citet{lilly13} predicted an un-evolving FPZ unless the parameters governing
inflow/outflow/star formation evolve with redshift;
our approach suggests that indeed these parameters evolve.
\citet{yates12} study the evolution of the MZR from $z\sim 3$ to $z\simeq 0$
by comparing the semi-analytic models of \citet{guo11} with an SDSS-selected
sample similar to SDSS10.
However, their models were unable to predict the observed decrease of metallicity
for a fixed \mstar, and they concluded that ``chemical enrichment in the model
galaxies proceeded too rapidly at early times". 

Here we have shown that to correctly reproduce the coevolution of SFR and O/H, 
i.e., correctly fit the observations, that the {\it model
describing the co-dependence of \mstar, SFR, and metallicity must change with redshift}.
All three FPZ($z$) fits to both the MEGA dataset and the simulated galaxies 
as described above successfully
predict the O/H trends with redshift to within $0.03-0.05$\,dex.
Although the three formulations are apparently describing different physical
manifestations of the coevolution of metallicity and SFR,
they are generally painting a similar picture,
namely an increase of gas content with redshift.
%Below, we examine each of them in turn.

\subsection{Redshift variation of the model}
\label{sec:discussion_z}

According to Table \ref{tab:models},
the scaling of the FPZ($z$) model with \epsf\ suggests that: 
\begin{equation}
%\epsilon_*(z)\,\approx\,\epsilon_*(0)\,({\rm M}_{\rm star}/10^{9.5})^{-0.5}\,(1+z)^{-0.4} 
% 4/7/2016
\epsilon_*(z)\,\approx\,\epsilon_*(0)\,({\rm M}_{\rm star}/10^{9.5})^{-0.5}\,(1+z)^{-0.2} 
\label{eqn:epsilon_z}
\end{equation}
The above $\gamma$(\epsf), $\delta$(\epsf) coefficients are approximate, taken 
as a rough average of the (PP04N2) simulated and MEGA datasets.
Eqn. (\ref{eqn:epsilon_z}) would imply that at a given \mstar, \epsfinv, roughly the product of gas depletion time \taud\ and SF efficiency
\epff, is increasing slightly with redshift (roughly 30\% larger at $z\sim3$).
Assuming that \epff\ remains constant,
this would be contrary
to observational evidence that suggests that molecular depletion times
are shorter at higher redshift \citep[e.g.,][]{genzel10,genzel15,silverman15}.
Galaxies are also more gas-rich at higher redshift, with as much as
40$-$60\% of their dynamical mass in \htwo\ \citep[e.g.,][]{tacconi10,daddi10},
a fraction $\sim$5$-$10 times higher than typical spiral disks at $z\simeq 0$.
However, as discussed in Sect. \ref{sec:theory},
depletion times are defined
%SF efficiency (SFE) is defined 
observationally as the
%ratio of SFR to observed gas mass, thus proportional to the inverse
ratio of observed gas mass to SFR. 
%of gas depletion times and 
%unlike \epff, not a dimensionless quantity.
Our models do not know about {\it observed} gas mass; they only consider
the gas that is enriching the ISM with metals, i.e., the gas that is
effectively converted into stars.
Thus, of necessity, our models cannot distinguish between the observationally
defined \taud\ and the product of \taud\ and \epff, or equivalently
the ratio of \tff\ and \epff.
Given that \epsfinv\,=\,\mg(observed)\,\epff/$\psi$ (see Eqn. (\ref{eqn:epsilon}))
it is likely that the \epsf\ FPZ model scaling is 
pointing to increasing gas content with $z$,
rather than longer depletion times.

The FPZ($z$) formulation with the accretion and wind coefficients
is consistent with this.
As shown in Table \ref{tab:models},
the scaling of the model with $a$ and $w$:
% 4/7/2016 no change with respect to submitted version
\begin{eqnarray}
a(z)&\approx&a(0)\,(1+z)^{0.7}, \nonumber \\
w(z)&\approx&w(0)\,(1+z)^{0.0}
\label{eqn:aw_z}
\end{eqnarray}
indicates that the mass loading relative to SFR in the galactic outflows powered by star formation
is roughly the same at high $z$ as at $z\simeq 0$.
In contrast, the mass loading in accretion is significantly increased,
almost 3 times higher at $z\sim 3.5$ than locally.
Thus, to explain the drop in metallicity at high redshift,
more (pristine) gas is needed, 
in the model acquired through accretion.
Our result is roughly consistent with \citet{papovich11} who find
that gas fractions increase roughly as $(1+z)^{0.9}$
from $z\sim 3$ to $z\sim 8$, and after this ``accretion epoch'',
at lower redshifts both accretion and SFR are reduced.

A similar indication is given by the FPZ($z$) model based on 
the redshift variation of $\mu$, equal to the ratio of \mg\ and the initial
gas mass (see Eqn. (\ref{eqn:X})).
The coefficients in Table \ref{tab:models} indicate that:
% 4/7/2016
\begin{equation}
%\mu(z)\,\approx\,\mu(0)\,\frac{1}{({\rm M}_{\rm star}/10^{9.5})^{-0.4}\,(1+z)^{-0.3}}\quad .
\mu(z)\,\approx\,\mu(0)\,\frac{1}{({\rm M}_{\rm star}/10^{9.5})^{-0.4}\,(1+z)^{-0.15}}\quad .
\label{eqn:mu_z}
\end{equation}
Both the \mstar\ and redshift dependencies are similar to (although slightly smaller than)
those for the \epsf\ formulation; 
this is not unexpected given that $\mu\,=\,M_g/M_{g0}\,\propto \epsilon_*^{-1}$.
The $\mu$ approach is perhaps more direct, but is essentially
indicating, as above, that more gas is needed with increasing redshift,
in order to form stars at the necessary levels, recycle the ISM,
and achieve the observed reduced metallicities.

%The increase in gas content with redshift roughly $\propto (1+z)^{0.3-0.4}$
The increase in gas content with redshift roughly $\propto (1+z)^{0.2}$
given by Eqns. (\ref{eqn:epsilon_z}) and
(\ref{eqn:mu_z}) is similar to (the inverse of) that found observationally 
for molecular depletion times \taud\ of main-sequence galaxies by \citet{genzel15}.
Since, as discussed above, our models are unable to separate
gas content from \taud\ and \epff, this is an encouraging consistency.
Moreover, for massive galaxies with \mstar\,$\approx 10^{11}$\,\msun,
%we would predict gas fractions at $z\sim 2$ roughly 6 times higher
we would predict gas fractions 7 times higher
at $z\sim 2$ 
than at $z\simeq 0$, roughly consistent with the CO observations by \citet{geach11},
and only slightly lower then the increase of a factor of 10 found by
\citet{bothwell13b} for luminous sub-millimeter galaxies.

Nevertheless, a limitation of our model is that it does not 
explicitly distinguish
between accreted gas, presumably \hi, and the gas which forms stars
that, at these high SFRs, must be molecular.
Indeed, the changes with redshift of relative gas content predicted by the 
FPZ($z$) model are slightly lower than the results derived observationally
for \htwo;
\citet{tacconi13} and \citet{saintonge13} find that, as $z$ goes from 2 to 1,
gas fractions decrease by a factor of $\sim$1.4, while we
would predict a change of only $\sim$1.1-1.2 (for a galaxy of fixed stellar mass).
The gas content of our model comprises not only the molecular
component but also \hi;
although it is not yet possible to observe \hi\ at high $z$,
it is likely that \hi\ content shows a smaller redshift variation
than \htwo\ 
\citep[e.g.,][]{lagos11,popping12,lagos16}.
Thus, total gas fractions are expected to increase less rapidly with
redshift than \htwo\ alone.
More detailed comparisons with observations will require observations
of atomic gas to cosmological redshifts which should be possible
with the Square Kilometer Array (SKA).

\subsection{Redshift variation of the stellar mass scaling}
\label{sec:discussion_mstar}

It is well established that at a given redshift
the gas fraction in massive galaxies is lower
than in less massive ones
\citep[e.g.,][]{saintonge11,popping12,huang12,boselli14,bothwell14,popping15}.
Moreover, as discussed above, 
it is clear that gas fraction increases with redshift. 
The FPZ($z$) model with the \epsf\ and $\mu$ approaches
also predicts that the amount of increase in gas content with redshift
varies with stellar mass.
In particular,
the \mstar\ scaling of the FPZ($z$) models
(see Eqns. (\ref{eqn:epsilon_z}) and (\ref{eqn:mu_z}))
suggests that the gas content of high-mass galaxies
relative to less massive ones {\it increases} with increasing redshift.
For a galaxy with \mstar\,=\,$10^{10}$\,\msun\ at $z\,=\,2$,
the growth in gas content is predicted to be $\sim 2.2$ times
that at $z\,=\,0$;
a more massive galaxy, \mstar\,=\,$10^{11}$\,\msun\ at the
same redshift the increase would be $\sim 7$ times the
gas content of a galaxy of the same mass at $z\,=\,0$. 
A less massive galaxy, with \mstar\,=\,$10^{9}$\,\msun, would
be expected to have roughly the same gas content at $z\,=\,2$
as at $z\,=\,0$.
Such a result would be consistent with ``downsizing'' in which
massive galaxies evolve more rapidly than lower-mass ones,
thus consuming their gas at earlier epochs
\citep[e.g.,][]{cowie96,delucia06,bundy06,thomas10}.
%\textcolor{purple}{
Specifically, \citet{thomas10} found that SF activity in
galaxies with \mstar\ $\sim 10^{11}-4\times10^{11}$\,\msun\ peaked at increasingly
higher redshifts, $z\sim 1.5-2$, well within the redshift and mass ranges
probed by the MEGA and simulated galaxies.
%}

Such a trend of gas fraction with stellar mass 
and redshift is also in accord with some observational estimates
of the mass variations of gas fraction with redshift,
either through indirect determinations of gas content by inverting gas-SFR
scaling relations \citep[e.g.,][]{popping12,popping15},
or with gas masses inferred from measured dust masses up to $z\sim 2.5$
after correcting for metallicity \citep{santini14}.
\citet{santini14} find that from $z\sim 2.5$ to $z \sim 1$,
the gas fraction in massive galaxies decreases more
sharply than in lower-mass ones;
less massive galaxies show a shallower decrease in the
gas content.

However, observations of \htwo\ alone are in disagreement
with this.
\citet{morokuma15} estimated redshift
variations of \htwo\ fraction through a compilation of CO observations 
and 
find that more massive galaxies show {\it less} evolution in \htwo\ fraction than
lower-mass ones, in direct contrast to our results. 
The same contradiction emerges in the study by \citet{dessauges15}
who studied five lensed galaxies and compared them with CO
observations from the literature.
\citet{genel14} Illustris simulations also find that less massive
galaxies at higher redshift have a larger change in gas fraction
with redshift.
Observationally, our result for galaxies with masses $\la 10^{10}$\,\msun\
is very difficult to verify given that virtually all observational studies
so far of gas content at high redshift are limited to
\mstar$\ga\ 10^{10}$\,\msun\ \citep[although see][who studied
lensed galaxies]{saintonge13,dessauges15}.
Nevertheless, the results of the FPZ($z$) models suggest that
massive galaxies at $z\sim 3$ relative to $z\simeq 0$ must have a higher gas
content in order to achieve the relatively lower metallicities.
Incorporating a multi-phase approach as in \citet{magrini12} might
help interpret the different gas components in our models.

One limitation of the FPZ model was discussed above, namely the inability
to distinguish between \htwo, \hi, and total gas content.
%There is at least one other possible limitation to the model, namely the lack
At least one other possible limitation to the model is the lack
of mass scaling of \epsf, or equivalently the SF timescale.
Such as omission can be justified for the SDSS10 galaxies because, as
discussed in Sect. \ref{sec:theory}, in typical
spiral disks both molecular depletion times \taud\ and \hi\
depletion time are roughly constant, $2-3$\,Gyr
\citep{bigiel11,catinella10,schiminovich10}.
Nevertheless, this could be a problem for the MEGA dataset
because of its highly star-forming nature, and large variation
in \mstar.
It could also be troublesome for the high-$z$ main-sequence
simulated galaxies because
of their relatively higher SFRs compared to local ones.
It is possible that the \mstar\ scaling we find for the \epsf\
and $\mu$ adaptations of the FPZ model to $z>0$ is a consequence 
of neglecting such a treatment at $z\,=\,0$.

%\textcolor{purple}{ 
Our treatment of trends of FPZ parameters with $z$ and \mstar\
as simple power laws is also highly over-simplified.
As found by \citet{lagos11,lagos14,morokuma15}, the redshift and \mstar\
dependence of gas-mass fractions is more complicated than this, as there
are inflections and slight curvatures in the behavior of both quantities.
Thus, our models are almost certainly not exact, but rather a 
simplified scaling aimed at a general representation of the trends
necessary to explain the coevolution of SFR and metallicity.
%}

%\textcolor{purple}{
Recently it has been proposed that scaling laws based on 
gas fraction, SFR, and \mstar\ 
are more fundamental than those based on metallicity
\citep[e.g.,][]{bothwell13a,santini14,lagos16,bothwell16}.
The physical basis of such a scaling is clear, since gas provides
the fuel for star formation and the reservoir of gas-phase metals.
However, the role of \hi\ vs. \htwo\ is still under debate, and the possibility
of verifying through observations the amount of total gas in \hi\ 
is currently not possible, although SKA will be able to help
shed light on this problem.
%}

\section{Summary}
\label{sec:conclusions}

%\begin{itemize}
In a companion paper, we constructed a new MEGA dataset for studying the coevolution
of metallicity and SFR;
the data are compiled from 19 different samples up to $z \simeq 3.7$,  
spanning a factor of $\sim 10^5$ in \mstar, $\sim 10^6$ in SFR, and almost two
orders of magnitude in O/H.
As a comparison sample, we include the SDSS10 galaxies studied by \citet{mannucci10}.

\noindent
\hangindent=0.05\linewidth
\hangafter=1
%\item 
$\bullet$\ \
Here we update the model of 
\citet{dayal13} for the $z\simeq 0$ behavior
of \mstar, SFR, and O/H and apply it to the SDSS10 and MEGA datasets;
the new model relies on the stellar yields from \citet{vincenzo16} and
\citet{nomoto13} assuming a \citet{chabrier03} IMF.
With only 5 free
parameters, the FPZ model at $z\simeq 0$ is able to reproduce the observed oxygen
abundance of the $\sim$80\,000 SDSS10 galaxies to within 0.05$-$0.06\,dex,
and of the $\sim 250\ z\simeq 0$ MEGA galaxies to within $0.24-0.26$\,dex.

\noindent
\hangindent=0.05\linewidth
\hangafter=1
%\item 
$\bullet$\ \ 
We have extended the FPZ($z\simeq 0$) model to higher redshift by
exploring three possibilities: 
(1) a redshift (and mass) scaling of SF timescale \epsf\ (Sect. \ref{sec:epsilon_z}); 
(2) a redshift scaling of accretion and galactic wind mass loading (Sect. \ref{sec:aw_z}); and
(3) a redshift (and mass) scaling of gas fraction (Sect. \ref{sec:mgas_z}).
These approaches give similar results, and are able to reproduce the 
observed metallicity trends in the MEGA sample to within $\sim 0.03-0.05$\,dex.

\noindent
\hangindent=0.05\linewidth
\hangafter=1
%\item 
$\bullet$\ \ 
The extension of the FPZ($z$) model allows us to quantify how gas accretion and outflow depend on
redshift. Although the specific mass loading of outflows does not change
measurably during the evolution, the accretion rate and gas content of galaxies
increase significantly with redshift. These two effects 
explain, either separately or possibly in tandem, 
the observed lower metal abundance of high-$z$ galaxies.

%\end{itemize}

\section*{Acknowledgments}
We thank Filippo Fraternali for interesting discussion and insightful comments.
This work has benefited from the DAVID network
({\it http://wiki.arcetri.astro.it/DAVID/WebHome}) for fostering a fruitful collaborative environment.
PD gladly acknowledges funding from the EU COFUND Rosalind Franklin program, and
LH from a national research grant, PRIN-INAF/2012.

\label{lastpage}

\end{document}